\documentclass[12pt]{article}
\usepackage{graphicx}
\usepackage[margin=1.25in]{geometry}
\usepackage{color}
\usepackage{url}
\usepackage[colorlinks = true,
            linkcolor = blue,
            urlcolor  = blue,
            citecolor = blue,
            anchorcolor = blue]{hyperref}

\newcommand\pubnumber{ILC--NOTE--2016--067 \\DESY 16--145, 
  IPMU16-0108\\  KEK Preprint 2016--9, 
LAL 16--185\\ MPP--2016--174,
 SLAC--PUB--16751 }
\newcommand\pubdate{July, 2016}

\def\SLAC{SLAC,
    Stanford University, Menlo Park, CA 94025, USA}

\def\Title#1{\begin{center} {\Large #1 } \end{center}}
\def\Author#1{\begin{center}{ \sc #1} \end{center}}

\newenvironment{Abstract}{\begin{quotation} \begin{center}
                       ABSTRACT
     \end{center}\bigskip  }{\end{quotation}}

\def\Acknowledgements{\bigskip  \bigskip \begin{center} 
             \bf ACKNOWLEDGEMENTS \end{center}}
\newcommand\pubblock{\rightline{\begin{tabular}{l} \pubnumber\\  \\
         \pubdate \end{tabular}}}
\def\beq{\begin{equation}}
\def\eeq#1{\label{#1}\end{equation}}
\def\eeqn{\end{equation}}

\newenvironment{Eqnarray}%
   {\arraycolsep 0.14em\begin{eqnarray}}{\end{eqnarray}}
\def\beqa{\begin{Eqnarray}}
\def\eeqa#1{\label{#1}\end{Eqnarray}}
\def\eeqan{\end{Eqnarray}}
\def\CR{\nonumber \\ }

\def\leqn#1{(\ref{#1})}

\let\bar=\overbar

\def\lsim{\mathrel{\raise.3ex\hbox{$<$\kern-.75em\lower1ex\hbox{$\sim$}}}}
\def\gsim{\mathrel{\raise.3ex\hbox{$>$\kern-.75em\lower1ex\hbox{$\sim$}}}}

\def\L{{\cal L}}

\def\tr{{\mbox{\rm tr}}}
\def\half{\frac{1}{2}}

\def\third{\frac{1}{3}}

\def\del{\partial}
\def\Dslash{\not{\hbox{\kern-4pt $D$}}}
\def\dslash{\not{\hbox{\kern-2pt $\del$}}}

\def\ee{e^+e^-}
\def\gmgm{\gamma\gamma}

\def\mz{m_Z}

\def\msb{{\bar{\scriptsize M \kern -1pt S}}}

\def\drb{{\bar{\scriptsize D \kern -1pt R}}}

\makeatletter
\def\section{\@startsection{section}{0}{\z@}{5.5ex plus .5ex minus
 1.5ex}{2.3ex plus .2ex}{\large\bf}}
\def\subsection{\@startsection{subsection}{1}{\z@}{3.5ex plus .5ex minus
 1.5ex}{1.3ex plus .2ex}{\normalsize\bf}}
\def\subsubsection{\@startsection{subsubsection}{2}{\z@}{-3.5ex plus
-1ex minus  -.2ex}{2.3ex plus .2ex}{\normalsize\sl}}

\renewcommand{\@makecaption}[2]{%
   \vskip 10pt
   \setbox\@tempboxa\hbox{\small #1: #2}
   \ifdim \wd\@tempboxa >\hsize     % IF longer than one line:
       \small #1: #2\par          %   THEN set as ordinary paragraph.
     \else                        %   ELSE  center.
       \hbox to\hsize{\hfil\box\@tempboxa\hfil}
   \fi}

 \def\citenum#1{{\def\@cite##1##2{##1}\cite{#1}}}
\newcount\@tempcntc
\def\@citex[#1]#2{\if@filesw\immediate\write\@auxout{\string\citation{#2}}\fi
  \@tempcnta\z@\@tempcntb\m@ne\def\@citea{}\@cite{\@for\@citeb:=#2\do
    {\@ifundefined
       {b@\@citeb}{\@citeo\@tempcntb\m@ne\@citea\def\@citea{,}{\bf ?}\@warning
       {Citation `\@citeb' on page \thepage \space undefined}}%
    {\setbox\z@\hbox{\global\@tempcntc0\csname b@\@citeb\endcsname\relax}%
     \ifnum\@tempcntc=\z@ \@citeo\@tempcntb\m@ne
       \@citea\def\@citea{,}\hbox{\csname b@\@citeb\endcsname}%
     \else
      \advance\@tempcntb\@ne
      \ifnum\@tempcntb=\@tempcntc
      \else\advance\@tempcntb\m@ne\@citeo
      \@tempcnta\@tempcntc\@tempcntb\@tempcntc\fi\fi}}\@citeo}{#1}}
\def\@citeo{\ifnum\@tempcnta>\@tempcntb\else\@citea\def\@citea{,}%
  \ifnum\@tempcnta=\@tempcntb\the\@tempcnta\else
  {\advance\@tempcnta\@ne\ifnum\@tempcnta=\@tempcntb \else\def\@citea{--}\fi
    \advance\@tempcnta\m@ne\the\@tempcnta\@citea\the\@tempcntb}\fi\fi}
\makeatother

\def\kek{High Energy Accelerator Research Organization (KEK), Tsukuba,
  Ibaraki, JAPAN  }
\def\ifae{ICREA at IFAE, Univesitat Aut\'onoma de Barcelona, E-08193
  Bellaterra, SPAIN   }
\def\Tsinghua{Center for High Energy Physics, Tsinghua University, Beijing, CHINA}
\def\Toyama{Department of Physics, University of Toyama, Toyama 930-8555, JAPAN}
\def\Seoul{Department of Physics and Astronomy, Seoul National
  University, Seoul 151-747,  \\ \hskip 0.4in   KOREA}
\def\DESY{DESY, Notkestrasse 85, 22607 Hamburg, GERMANY}
\def\Cornell{Laboratory for Elementary Particle Physics, Cornell
  University, Ithaca, NY 14853, \\ \hskip 0.4in USA  }
\def\Orsay{LAL, Centre Scientifique d'Orsay, Universit\'e Paris-Sud, 
  F-91898 Orsay CEDEX, \\ \hskip 0.4in FRANCE }
\def\OrsayTH{LPT, Universit\'e Paris-Sud, 91405 Orsay, FRANCE}
\def\MPP{Max-Planck-Institut f\"ür Physik, F\"öhringer Ring 6, 80805 Munich, GERMANY}
\def\Tokyo{ICEPP, University of Tokyo, Hongo, Bunkyo-ku, Tokyo,
  113-0033, JAPAN}
\def\UTA{Department of Physics, University of Texas, Arlington, TX 76019, USA}
\def\Michigan{Michigan Center for Theoretical Physics, University of
  Michigan, Ann Arbor, \\ \hskip 0.4in MI 48109, USA}
\def\Berkeley{Department of Physics, University of California,
  Berkeley, CA 94720, USA}
\def\LBL{Theoretical Physics Group, Lawrence Berkeley National
  Laboratory, Berkeley, \\  \hskip 0.4in   CA 94720, USA}
\def\IPMU{Kavli Institute for the Physics and Mathematics of the
  Universe, \\  \hskip 0.4in   University of Tokyo, Kashiwa 277-8583, JAPAN}
\def\Tohoku{Department of Physics, Tohoku University, Sendai, Miyagi 980-8578, JAPAN}
\def\TokyoTh{Department of Physics, University of Tokyo, Tokyo
  113-0033, JAPAN}
\def\Valencia{ IFIC (UVEG/CSIC), Edificios de Investigacion,
c./ Catedratico Jose Beltran 2, E-46980 Paterna, Valencia, SPAIN}
\def\KIAS{Quantum Universe Center, KIAS, Seoul 02455, KOREA}

\begin{document}
\begin{titlepage}
\pubblock

\vfill
\Title{Implications of the 750 GeV $\gamma\gamma$ Resonance  as a Case
  Study for the International  Linear Collider}
\vfill
\Author{LCC Physics Working Group}
\bigskip
\Author{Keisuke Fujii$^1$, Christophe
Grojean$^{2,3}$ Michael E. Peskin$^4$(conveners); Tim
Barklow$^4$, Yuanning Gao$^5$,
Shinya Kanemura$^6$, Hyungdo Kim$^7$, Jenny List$^2$,
Mihoko Nojiri$^{1,8}$, Maxim
Perelstein$^9$, Roman P\"oschl$^{10}$,  J\"urgen Reuter$^2$, Frank
Simon$^{11}$, 
Tomohiko Tanabe$^{12}$, Jaehoon Yu$^{13}$,
James D. Wells$^{14}$;  Adam Falkowski$^{15}$, Shigeki
Matsumoto$^{8}$,\\ Takeo Moroi$^{16}$, 
Francois Richard$^{10}$, 
Junping Tian$^{12}$,
Marcel Vos$^{17}$, Hiroshi Yokoya$^{18}$;  Hitoshi
Murayama$^{8,19,20}$, 
Hitoshi Yamamoto$^{21}$}

\vfill
\begin{Abstract}
If the $\gamma\gamma$ resonance at 750 GeV suggested by 2015 LHC data
turns out to be a real effect, what are the implications for the
physics case and upgrade path of the International Linear Collider?
Whether or not the resonance is confirmed, this question provides an interesting
case study testing the robustness of  the ILC physics case.
In this note, we address this question with two points:  (1) Almost all
models proposed for the new 750 GeV particle require additional  new
particles with electroweak couplings.  The key
elements of the 500 GeV ILC physics program---precision
measurements 
of the Higgs boson, the top quark, and 4-fermion
interactions---will powerfully discriminate among these models.  This
information will be important in conjunction with new LHC data, or alone,
if the new particles accompanying the 750
GeV resonance are beyond the
mass reach of the LHC.  (2)  Over a longer term, the energy upgrade of the ILC to
1~TeV already discussed in the ILC TDR 
will enable experiments in $\gamma\gamma$ and $\ee$ collisions
to  directly produce and study the 750~GeV particle from these unique
initial states.
\end{Abstract}

\vfill

\vfill

% \begin{small} 
% Work supported in part by the US Department of Energy,
%                      contract DE--AC02--76SF00515.
% \end{small}

\newpage
\mbox{\quad}
\vfill

\begin{raggedright}
\noindent $^1$ \kek \\
$^2$  \DESY \\
$^3$  \ifae \\
$^4$ \SLAC\\
$^5$ \Tsinghua \\
$^6$  \Toyama\\
$^7$ \Seoul \\
$^8$ \IPMU \\ 
$^9$ \Cornell\\
$^{10}$  \Orsay\\
$^{11}$ \MPP \\
$^{12}$  \Tokyo\\
$^{13}$ \UTA\\
$^{14}$  \Michigan \\
$^{15}$  \OrsayTH\\
$^{16}$  \TokyoTh \\
$^{17}$  \Valencia \\
$^{18}$ \KIAS\\
$^{19}$ \Berkeley \\
$^{20}$ \LBL  \\
$^{21}$  \Tohoku \\
\end{raggedright} 

\vfill

\newpage
\mbox{\quad}

\vfill
\tableofcontents

\vfill

\newpage

\def\thefootnote{\fnsymbol{footnote}}
\setcounter{footnote}{0}
\newpage
\mbox{\quad}

\end{titlepage}

\section{Introduction}

There are many arguments that there exist new interactions of
physics beyond the currently defined Standard Model (SM)  of particle
physics.   The main features of the observed universe---the presence
of a small, nonzero dark energy, the presence of cold dark matter at a
total mass 5 times that of atomic matter, and the prevalence of
baryons over antibaryons---cannot be accounted for by the SM.
However, today there are no unambiguous signals that
point to a strategy for exploring for physics beyond the SM.  We fall
back, then, on the obvious strategies of searching for new particles
at high energy and measuring the properties of known particles with 
increasing precision.

If we could have more specific information about the nature of the new
physics, this might help us define better the goals of a program of
 future accelerator
experiments.  It is interesting to look even at suggested anomalies in
the data in this light.  We should ask: If the suggested effect is
confirmed, what path to new physics would be suggested, and how can
proposed future accelerator facilities explore this path?

In December 2015, the ATLAS and CMS collaborations presented very
preliminary evidence for a resonance at a  mass of about 750~GeV,
created in $pp$ collisions and decaying to
$\gamma\gamma$~\cite{ATLAS,CMS}.    These results, and the results
of the searches for  $\gamma\gamma$ resonances in the 8~TeV data, were 
recently updated~\cite{ATLAS2,CMS2}. 
At this time, the evidence for this resonance is hardly persuasive.
ATLAS quotes a local significance of the effect in the 13~TeV data at
about 3.7~$\sigma$ and a
global significance of 2.0~$\sigma$, and a local significance in the 8
TeV data of 1.9~$\sigma$.  CMS reports a peak with local
significance 2.8~$\sigma$ and a small compatible effect in the 8~TeV
data.
  This would be unremarkable except that
its location seems to coincide with the peak location  found by ATLAS.
Clearly, more data is needed to confirm this observation.  

The purpose of this note is to gather information on the question:  If
the resonance suggested by the LHC data is real, what would the
implications be for the program of the International Linear Collider?   
Will the ILC be able to
shed light on this resonance or on accompanying new physics?  For
definiteness, we  refer to the  resonance from here on as $\Phi$.

Since the ILC is designed for an energy of 500 GeV, it will not be
able to produce the $\Phi$ directly.  The ILC TDR foresees an upgrade of
the accelerator to a center of mass energy of 1000 GeV, and this
machine will be able to produce $\Phi$ either from the $\ee$ or from the 
$\gmgm$ initial state.  One might discuss a $\gamma\gamma$
collider optimized for $\Phi$ production, and we will give parameters
below, 
following~\cite{Badelek:2001xb,Ito:2016zkz,Djouadi:2016eyy,ZGHe,Richard:2016nhm}.

However, a discussion of the implications of the $\Phi$ resonance
should not focus exclusively on direct observation of the $\Phi$.  As
we will discuss, in most of the models that have been proposed for the
identity of the $\Phi$, this resonance is one of many new particles
that must be introduced.  In most models, other required new particles
should be discovered at the LHC.   It might turn out that the $\Phi$
is a relatively minor player in a new sector of physics that the LHC
will begin to uncover in the next few years.  For this reason, it is 
premature to discuss a new accelerator intended specifically to target
the $\Phi$ or any other new particle that turns up in the early 13~TeV
LHC data.

The relation of the ILC to the $\Phi$, however, is quite different.
In presentations of the physics goals of the ILC (for example,
\cite{TDR,Case,Barklow:2015tja}), it is always emphasized that the ILC at 500~GeV offers 
techniques for observing effects of new physics that are orthogonal to
those of direct particle search at the LHC and are 
sensitive to a broad range of new physics models.   These include 
precision measurements of the 
couplings of the Higgs boson, the electroweak couplings of the top
quark, the 3-boson couplings of the $W$ and $Z$ bosons, and
interactions
mediating fermion-fermion scattering.     If the observation of the
$\Phi$ is confirmed, and even if the LHC discovers further additional
new particles related to the $\Phi$, 
these techniques will be indispensable to
discriminate possible explanations of the resonance and to demonstrate
the presence of further states beyond the reach of the LHC.
If the $\Phi$ is not confirmed, these techniques will still provide a means to
search for physics beyond the Standard Model and might provide the
first discovery of new physics.

The models that have been suggested for the identity of the $\Phi$
highlight these capabilities by predicting substantial effects in the
ILC precision experiments.   As such, they provide a worked example of
the impact of the ILC precision measurements.  The study of this
example gives insight,
whether or not the $\Phi$ turns out to be a real signal.   

The structure of this note is as follows:  In Section 2, we review
basic properties of the $\Phi$ resonance necessary for our discussion.
In Section 3, the heart of the paper, we present the importance of
precision measurements on Higgs, top, and 4-fermion interactions for a
variety of specific models that have been proposed for the $\Phi$.  At
the same time, we should not ignore the capabilities of an
energy-upgraded ILC to directly produce the $\Phi$ and its partners. 
In Section 4, we
discuss the observation of the $\Phi$ at the ILC, upgraded to 1000~GeV,
in $\gmgm$ and $\ee$ collisions, and the possibility of a
$\gamma\gamma$ collider optimized for the $\Phi$ resonance.

\section{Properties of the $\Phi$}

To orient this discussion, it is useful to recall what we 
know about the $\Phi$, assuming that it is not a statistical fluctuation.  

Since the  $\Phi$ decays to two photons, it must be a color singlet
state and cannot have spin 1. In most of our discussion, we will
assume that the $\Phi$ is a spin 0 particle, either scalar or
pseudoscalar.   It is also possible that the $\Phi$ has spin 2; we
will discuss a model of this type in 
Section~3.6.    It has also been proposed that the $\Phi$
enhancement is a kinematic endpoint in the decay of a particle with
mass above 1.5~TeV~\cite{Kim:2015ron,Kong,Knapen:2015dap}.   To keep this discussion
finite, we will concentrate here on the hypothesis that $\Phi$ is a
spin 0 or spin 2 resonance, though some of the models discussed (e.g., in
Section 3.4) would also
     be compatible with the kinematic endpoint hypothesis.

The $\Phi$ is seen in the 13 TeV LHC data but is much less apparent in the 8~TeV
data. The ATLAS and CMS datasets are about 20~fb$^{-1}$ at 8~TeV and
3.2~fb$^{-1}$ (ATLAS), 2.6~fb$^{-1}$ (CMS) at
13~TeV, so, even though parton luminosities are higher at 13~TeV,
there is tension between the two results.
To minimize this tension, we   prefer models in which the parton luminosity has a substantial
increase from 8 to 13 TeV~\cite{Knapen:2015dap,CERN,Buttazzo:2015txu,Gupta:2015zzs}. 
 Let $r_{p\bar p}$ be the ratio of parton
luminosities
between 13 and 8 TeV.  Using the NNPDF2.3QED distribution
functions~\cite{NNPDF}, we find
\beq
\begin{tabular}{l|ccccc}
 & $gg$  &  $b\bar b$  &  $d \bar d$ &   $u\bar u$ &
 $\gamma\gamma$ \\ \hline
$r_{p\bar p}$ &  4.8 &  5.7  &  2.7 & 2.6 &  1.9   
\end{tabular}
\eeq{rtable}
Production from $gg$ gives the best combination of high parton
luminosity and high ratio, while if the  production is  from
$\gamma\gamma$, it is very difficult to reconcile the results from 8
and 13~TeV.    From here on, we will assume that the $\Phi$     is  dominantly
produced through $gg\to \Phi$.

The excess of about 20 events between the two experiments, 
collected with 50\% efficiency, in 5.8~fb$^{-1}$ yields a cross
section of about  7~fb.  However, this should be combined with constraints
from the small event rate at  8 TeV.   In this paper, we will take a
conservative reference value of
the cross section:
\beq
       \sigma(pp \to \Phi \to \gamma\gamma) =   5\ \mbox{fb} \ .
\eeq{refcross}

The cross section for gluon fusion to a spin $J$ resonance is given in
leading order by 
\beq
    \sigma(gg \to \Phi \to \gamma\gamma)  =  {\pi^2\over
      8}(2J+1) {\Gamma(\Phi\to gg)\over m_\Phi} \delta(\hat s - m_\Phi^2)
    \cdot {\Gamma(\Phi\to \gamma\gamma)\over \Gamma(\Phi) } \ ,
\eeq{csformula}
to be integrated over parton distributions. We can discuss this in
more detail for the case of a scalar resonance.  Higher order QCD corrections lead
to a $K$-factor for production of 2.8~\cite{Zurich}, but on the other
hand the QCD correction to the partial width  $\Gamma(\Phi\to gg)$ is
a factor of 2.0~\cite{Steinhauser}.
Dividing \leqn{refcross} by  the $gg$ parton
luminosity and the remaining factor of 1.4, we find that this reference
cross section implies
\beq
       {\Gamma(\Phi\to gg)\Gamma(\Phi\to \gamma\gamma)\over
         \Gamma(\Phi) }  =  0.5~\mbox{MeV}\ .
 \eeq{refrat}
In particular, if the dominant decay of the $\Phi$ is to $gg$, \leqn{refrat}
is equal to the partial width for $\Phi\to \gamma\gamma$,
\beq
     \Gamma(\Phi \to \gamma\gamma) =  0.5~\mbox{MeV} \ , 
\eeq{refGamma}
and
otherwise this value is a strict lower bound.

\begin{figure}
 \centering
 \includegraphics[width=0.5\hsize]{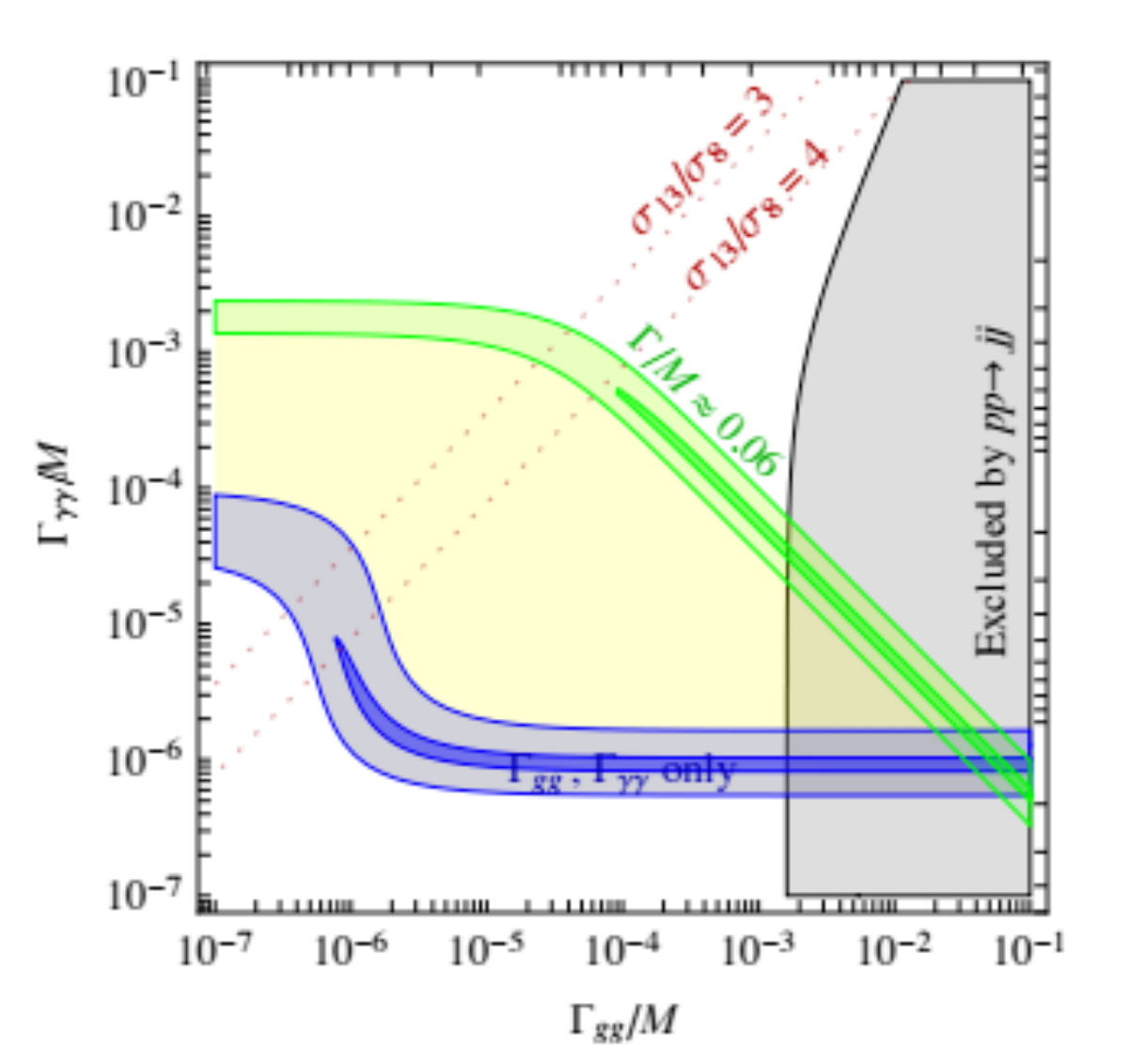}
  \caption{Region of the plane of $\Gamma(\Phi\to gg)$
    vs. $\Gamma(\Phi\to \gamma\gamma)$ suggested by the 2015 LHC data,
    from \cite{CERN}.   The purple band shows the expectation
if the total width of the $\Phi$ is dominated by the decays to $gg$
and $\gamma\gamma$.
The green band shows the expectation if the total width of the $\Phi$
is equal to 45~GeV, the value that gives the best fit to the 2015
ATLAS data.  The estimates in this paper assume
    $\Gamma(\Phi\to \gamma\gamma)/M = 0.7\times 10^{-6}$; thus, they
    are conservative.}
  \label{fig:CERN}
\end{figure}

The
prefered width to fit the ATLAS data is 45~GeV.   However, most models
of the $\Phi$ give much smaller total widths and also $BR(\Phi\to
\gamma\gamma) \sim 10^{-3} - 10^{-2}$. The range of possible values
for $\Gamma(\Phi\to \gamma\gamma)$ and $\Gamma(\Phi\to gg)$ is
illustrated
in Fig.~\ref{fig:CERN}~\cite{CERN}.   The lower purple band represents
the situation in which  $gg$ and $\gamma\gamma$ decays dominate the 
total width of the $\Phi$.  The upper green band corresponds to the
situation
that the total width of the $\Phi$ is indeed 45~GeV.  In this paper,
we will adopt the conservative assumption that $\Gamma(\Phi\to
\gamma\gamma)$  is given by \leqn{refGamma}.   The possibility that the
2-photon width of the $\Phi$ could be as large as 1~GeV is interesting
for ILC, but, in our opinion, this is not likely.   The preferred large
width could also result from a model in which the $\Phi$ is actually
formed as the sum of 
two nearby resonances. We will discuss this possibility in Section 3.2.

A useful phenomenological description of the coupling of $\Phi$ to
$gg$ and $\gamma\gamma$ is given by the effective Lagrangian
\beq
  {\cal L} = {\alpha_s\over 4} A_3\, \Phi \,G_{\mu\nu} G^{\mu\nu} 
+  {\alpha_w\over 4} A_2\, \Phi \,W_{\mu\nu} W^{\mu\nu}
 +  {\alpha^\prime\over 4} A_1 \, \Phi \, B_{\mu\nu} B^{\mu\nu}  \ ,
\eeq{effL}
where $G$, $W$, $B$ are the field strengths of the $SU(3)$, $SU(2)$,
$U(1)$ gauge fields of the SM and $\alpha_s$, $\alpha_w$, $\alpha'$
are the corresponding coupling constants.  We have written this
Lagrangian for a scalar $\Phi$; for a pseudoscalar, substitute  $F
\tilde F$ for $F^2$.    The constants $A_i$ have the dimensions of
(mass)$^{-1}$ and indicate the scale of the new interactions that give
rise to the coupling.  

There is an important point to be emphasized here:  The gauge
symmetries of the Standard Model imply that any coupling of $\Phi$ to
$gg$ or $\gamma\gamma$ must be through a non-renormalizable operator
whose coefficient introduces a new mass scale.
This implies the presence of new physics at that scale.  This is the same argument by
which the Fermi interaction implies new physics---the electroweak
interaction---at
 the 100~GeV mass
scale.   Another example is given by the 
125~GeV Higgs boson, whose the coupling to
$gg$ has a form similar to \leqn{effL}, with 
\beq
       A_3 =  {1\over 3 \pi v} \ ,
\eeq{Aorigin}
where $v$ is the Higgs vacuum expectation value, 246~GeV. This
coupling is generated by the top quark loop diagram.

 In this
formalism,
the
tree-level expressions for the $\Phi$ partial widths are (for $m_\Phi \gg
m_W, m_Z $)
\beqa
   \Gamma(\Phi\to gg) &=& { \alpha_s^2\over 8\pi}  A_3^2 m_\Phi^3 \CR
   \Gamma(\Phi\to \gamma\gamma) &=& { \alpha^2\over 64\pi}  (A_2 +
   A_1)^2  m_\Phi^3 \CR
   \Gamma(\Phi\to \gamma Z) &=& { \alpha \alpha_w c_w^2 \over 32\pi} (A_2 -
   {s_w^2\over c_w^2} A_1)^2  m_\Phi^3 \CR
    \Gamma(\Phi\to Z Z) &=& { \alpha_w^2 c_w^4 \over 64\pi} (A_2+
   {s_w^4\over c_w^4} A_1)^2  m_\Phi^3 \CR
   \Gamma(\Phi\to W^+W^-) &=& { \alpha_w^2\over 32\pi}  A_2^2 m_\Phi^3
   \ ,
\eeqa{allGammas}
where $(c_w,s_w) = (\cos\theta_w, \sin\theta_w)$.   The value
\leqn{refGamma} implies
\beq
   (A_2 + A_1) \sim  1/(500~\mbox{GeV}) \ .
\eeq{Avals}

The structure of \leqn{allGammas} implies that the $\Phi$ must decay
either to $\gamma Z$ or to $ZZ$.  Most likely, it decays to both
channels, with branching ratios comparable to the rate to
$\gamma\gamma$.  Because the $Z$ is most convincingly seen at the LHC
in its decay to $\ell^+\ell^-$ with a branching ratio of 7\%, it is
consistent with current data that these decays have not yet been
seen.  The fact that no resonance at 750 GeV has been observed in $WW$
and 
$ZZ$ at the LHC implies a relatively weak 
 limit $A_2/A_1 < 12$~\cite{Chala:2016mdz}.   If the $\Phi$ is
 confirmed, we would expect that its $\gamma Z$, $ZZ$, and $W^+W^-$
 decay modes would be observed at the LHC, and that these observations
 would clarify the effective Lagrangian \leqn{effL}.

Another piece of the evidence on  $\Phi$ comes from the fact that it
is not observed in other possible decay channels.  Limits on $\Phi$
production at 8~TeV are collected
 in~\cite{Knapen:2015dap,CERN,Buttazzo:2015txu,Gupta:2015zzs,Low:2015qep}.
A particularly useful summary can be found in Table~1 
of~\cite{CERN}.   
  A result of
particular significance here is that the $\Phi$ is not observed as
a resonance in Drell-Yan production ($\ee$ or $\mu^+\mu^-$), with a
cross section upper bound comparable to  the rate
\leqn{refcross} for production in $\gamma\gamma$~\cite{dielectron}.  
A spin 0 $\Phi$
would not be expected to decay to light leptons, but this is possible
in principle if the $\Phi$ has spin 2.  If $BR(\Phi\to \ee)$ has 
a value not far below the current upper bound, then the $\Phi$ could
appear as a prominent resonance in $\ee$ at 750~GeV.  We will discuss
the significance of the $\Phi ee$ coupling in Section 3.6.
Similarly, the $\Phi$ is not observed as a resonance in $t\bar t$ at a
level 
corresponding to 
$BR(\Phi\to t\bar t)/BR(\Phi\to \gamma\gamma) < 450$~\cite{ditop}.  
 This can be a significant
constraint for models that predict that $t\bar t$ is the dominant
decay mode of the $\Phi$. 

Beyond the question of the observable properties of the $\Phi$ lie
important physics issues.  In principle, the $\Phi$ could live in its
own sector of particles, completely disjoint from the known Higgs
boson.  But, it is highly suggestive that the $\Phi$ is somehow related
to the 125~GeV Higgs boson, or to other particles of the Standard
Model.  Thus, it is compelling to ask: 
 Does the $\Phi$ resonance  shed light on the origin of electroweak symmetry
breaking, or on other mysteries associated with the TeV energy scale?

If this is the question, measurements at the ILC are likely to give  an
important part of the answer.    Through its program of
precision measurements, the ILC
gives sensitivity to a very wide variety of new particles  with masses
below a few TeV.  In particular, these measurements  can discover effects from
almost any new particle that affects the 125~GeV Higgs boson or the
top quark.  Thus, the ILC will allow us to ask directly whether there
is a connection between the new and mysterious $\Phi$ resonance and
the more familiar particles at the 100 GeV mass scale.

\section{Imprint of the $\Phi$ on precision observables}

In this section, we will make more concrete how to obtain information
on the nature of the $\Phi$ from precision measurements that the ILC
will make available.   Many types of models
have been proposed for the $\Phi$. In almost all of these, the $\Phi$ is
accompanied by other types of new particles.   The proposed models
cover the whole range from weakly coupled models of extended Higgs
sectors to strongly coupled models in which the $\Phi$  is
composite.   Each model is built on a specific, unique relation between the
$\Phi$, the 125~GeV Higgs boson, and other new particles postulated in
the model.

There is no model-independent analysis that makes these relations
clear.  Instead, it is necessary to examine the models one by one,
understanding,  in each case, the characteristic structure  of the
model and the
impact predicted from this structure on known particles of the Standard Model.
In this section, we will describe the various classes of  models
proposed for the $\Phi$ and clarify these relations in each case.   We
will also point out, for each class of models, the specific effects that should
be visible in ILC measurements.  We will cite explicit realizations of
models of each class.   A more complete bibliography of the literature
on the $\Phi$, which now includes more than 400 theoretical papers,
can be found in~\cite{Strumia}. 

The literature on the $\Phi$ has shown that essentially every approach
to new physics at the TeV scale can accomodate the $\Phi$, either by
identifying the  $\Phi$ with a particle already present in the model
or by adding the $\Phi$ in a simple  way.   In models of the first
type, the parameter values needed to include the $\Phi$ are often quite
different from those anticipated prior to the $\Phi$ observation.  In
both cases, the new parameters often  seem  fine-tuned.  In the
discussion below, we simply accept these unusual or tuned parameters
as the price of describing the $\Phi$ successfully.

In most cases, the existing literature on the implications of
precision Higgs, top, and electroweak couplings already suggests
methods by which measurement of these couplings can distinguish models
of the various classes.   Typically, the parameter choices needed to
accomodate the $\Phi$ lead to effects in the precision measurements
that are larger than 
those implied by generic choices.

\subsection{Singlet coupling to vectorlike quarks and leptons}

\def\lsim{\stackrel{<}{_\sim}}
\def\gsim{\stackrel{>}{_\sim}}

The simplest models of the $\Phi$ build a minimal structure around the
effective Lagrangian \leqn{effL}.    In these models, the $\Phi$ is a
scalar or pseudoscalar 
Standard Model singlet.   We must couple this state to heavy particles
that can generate \leqn{effL} when they are integrated out.  To obtain
sufficiently large couplings $A_i$, these heavy particles must, in
most models, be fermions rather than scalars.  It is not possible to
use conventional 4th-generation quarks and leptons.   For example, any heavy quark that
obtains its full mass from the 125 GeV Higgs boson gives a
contribution to the $hgg$ amplitude equal to that of the top quark.  A
heavy
fourth-generation quark doublet would then  increase the $hgg$ couplings by
a factor 3 and the cross section $\sigma(gg\to h)$ by a factor 9.   To
generate \leqn{effL} without such large effects on the 125 GeV Higgs,
the heavy fermions should have the same $SU(2)\times U(1)$ quantum
numbers for their left- and right-handed components, so that they can
obtain mass without invoking the Higgs field expectation value.   Such
fermions are said to be ``vectorlike''.   In some models, the heavy
fermions obtain mass from the vacuum expectation value of the
field that gives rise to the
$\Phi$~\cite{Knapen:2015dap,CERN,Buttazzo:2015txu,Ellis:2015oso,Falkowski:2015swt,Berthier:2015vbb,Craig:2015lra,Dev:2015vjd,Dolan:2016eki,Cao:2016udb}.

To
generate couplings both to gluons and to photons, the heavy particles
must include states with both QCD 
and electroweak quantum numbers.
In the simplest model, both types of couplings are generated by a
heavy vectorlike quark.  However, it is also possible that the gluon
and electroweak
couplings are generated by different particles, 
and, in particular, that the couplings to the photon are generated by new heavy leptons.

As an example, consider adding to the Standard Model a vectorlike 
heavy quark  $Q$. This quark would couple to a
scalar $\Phi$ via a Yukawa coupling
\beq
{\cal L} \subset \ (m_\Psi + y\Phi )\bar{Q} Q\ . 
\eeq{L_VLQ}   
(If $\Phi$ is a pseudoscalar, substitute $\bar{\Psi} \gamma^5 \Psi$ in
the second term.)   The contributions to the $A_i$ in \leqn{effL} are
proportional to $y/m_\Psi$.  For 
example, introduce a color-triplet, weak-singlet vectorlike quark of
electric charge $5/3$.  A calculation similar to \leqn{Aorigin} gives
\beq
A_1={50\over 9} {y\over \pi m_Q} \ , \quad A_2=0\ , \quad A_3={1\over
  3} {y\over \pi m_Q} \ .
\eeq{As}
From the production cross section \leqn{refcross}, we infer $m_Q/y\sim
900$~GeV. Then $y$ would lie in the interval between $y\sim 0.7$, where
$Q$  should already have been discovered at the LHC, and $y \sim 4$,
at which the coupling is so large that perturbation theory no longer applies.  Then $Q$ 
can probably be discovered at the LHC with 3 ab$^{-1}$, except at the
largest allowed values of $y$. 
In a model with several  heavy vectorlike quarks, each contributes
according to \leqn{As}, and the estimated masses
are larger.

Another possible case is generated by adding a vectorlike color octet
fermion $G$ with zero electroweak quantum numbers and a vectorlike
lepton $E$ with $(I,Y) = (0,1)$. 
 In this case
 \beq
A_1=   {2 y_E\over \pi m_E} \ , \quad A_2=0\ , \quad A_3=2 {y_G\over
  \pi m_G} \ .
\eeq{Es}
In this case, the masses of $G$ and $E$ are not coupled in the
phenomenology.  The mass of $G$ can easily be above 3~TeV, beyond the
reach of the LHC.  The mass of 
$E$ should be below 1~TeV.   The LHC can discover the $E$ in some but
not all of this range.   The paper~\cite{Kumar:2015tna} studies 
many possible decay
schemes for the $E$ and computes expected limits from the 
High-Luminosity LHC 
ranging from 200~GeV to 500~GeV.

If $\Phi$ is a scalar, the theory must explain the absence of order-1
mixing between $\Phi$ and the SM doublet Higgs $H$, which is 
constrained by LHC measurements. To achieve this, one can postulate a $Z_2$
symmetry  under which $\Phi$ is odd and $H$ is even. However, the
Yukawa coupling  in \leqn{L_VLQ} breaks this symmetry, inducing the
doublet-singlet mixing through radiative corrections. This mixing in
turn induces shifts in the couplings of the 125 GeV Higgs boson to
gluons  and photons, typically in the $1-10$\% range accessible at the
ILC. Interestingly, the presence of the extra scalar in this model may
induce a strongly first-order electroweak phase transition, opening the
possibility of electroweak baryogenesis~\cite{Perelstein:2016cxy}. In
this scenario, 
Higgs couplings to photons and gluons must deviate by $5-10$\% from their SM values. 

Heavy vectorlike fermions are welcome in many theories of physics
beyond the Standard Model.  This is especially true for vectorlike
quarks.   In Little Higgs
theories~\cite{ArkaniHamed:2002qy,Schmaltz:2005ky,Perelstein:2005ka}
and certain versions of the Randall-Sundrum
model~\cite{Contino:2003ve}, vectorlike quark with charge $2/3$---top
quark partners---cancel the divergence in the Higgs boson mass coming
from top quark loops and thus are a crucial ingredient in solving the
hierarchy problem.  Top partners that play this role typically also
mix with the top quark, inducing shifts in the top quark couplings to
the weak gauge bosons. ILC measurements of the top width and its
couplings to the $Z$ are sensitive to these effects throughout the
natural region of 
parameter space in these models~\cite{Berger:2005ht}.  

Another possibility is that the $\Phi$ and the vectorlike
fermions arise within a grand unified theory.   For example, adding
extra ${\bf 5}+{\bf \bar {5} } $ representations to standard $SU(5)$
grand unification  adds  vectorlike fermions whose masses do not
depend on the Higgs field vacuum expectation value.  An interesting
and quite  predictive example is given in \cite{Hall:2015xds}.   This
is a supersymmetric model which introduces $ {\bf 5}+{\bf \bar {5}} $
multiplets that obtain mass from a the vacuum expectation value of a complex field $S$. 
The new  multiplets give rise to vectorlike quarks with $(I,Y) =
(0,-\third)$ and vectorlike leptons with $(I,Y) = (\half , -\half)$.
The assumption of unification at the GUT scale, plus renormalization
group running, fixes the masses of these particles to be in the ratio
2.5:1.   The overall mass scale is related to the size of the
coefficients in the effective Lagrangian \leqn{effL}.   The leptons
expected to have masses below 400~GeV, accessible for detailed study
at least at an energy-upgraded ILC.

\begin{figure}
\begin{center}
\includegraphics[width=0.99\hsize]{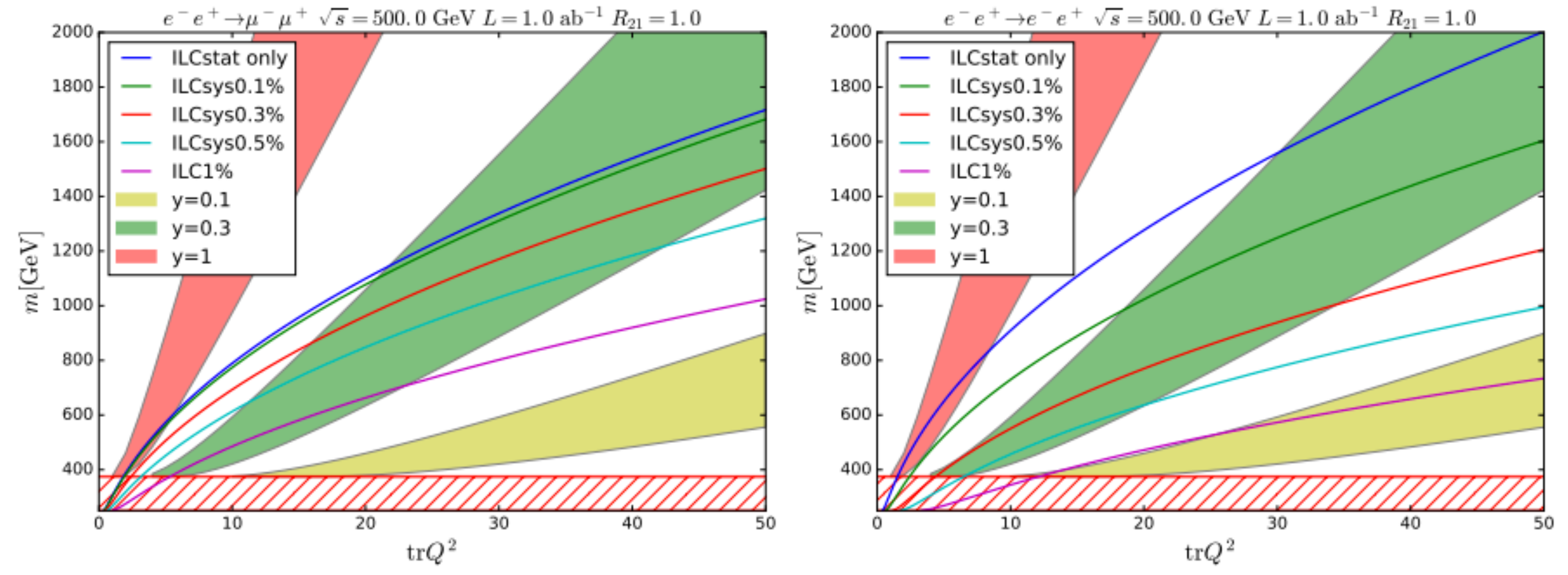}
\caption{Expected exclusion of vectorlike leptons by precision
  measurement of the cross sections for $e^+ e^- \to \mu^+ \mu^-$ and 
$e^+ e^- \to e^+ e^-$ at the ILC  at 500~GeV with 1000~ab$^{-1}$ of data, 
  from \cite{Bae:2016oey}, as a function of mass and multiplicity
  times squared charge.  The figure is shown for leptons with $I = Y =
  \half$.  The colored bands show the region of the plane that
  generates a $\Phi$ signal cross section of 3--10~fb, for the three  
values $y = 0.1, 0.3, 1$.}
\label{fig:Bae}
\end{center}
\end{figure}

The ILC can give  evidence for heavy fermions not only through direct
production but also, indirectly, through the effect of these particles
on the $U(1)$ and $SU(2)$ vacuum polarization amplitudes, which can be
extracted from precision measurement of 2-fermion scattering.
Figure~\ref{fig:Bae} gives an example~\cite{Bae:2016oey}.   In the
figure, the area below the curves can be excluded by ILC measurements
of  and $e^+ e^- \to \mu^+ \mu^-$ and $e^+ e^- \to e^+ e^-$  at
500~GeV with polarized electron beams $(P_{e^-},P_{e^+})=(-80\%,30\%)$
and 1~ab$^{-1}$ of data.   (The actual ILC run plan expects
4~ab$^{-1}$ of data at 500~GeV, divided among several  combinations of
beam polarization~\cite{Barklow:2015tja}.)
The colored bands show the regions of fermion mass and charge predicted for  a $\Phi$ signal
of 3-10~fb. It should be noted that models with $\tr[Q^2] > 6.7$
are inconsistent with standard grand unification, since some coupling
run to large values (or a Landau pole) below the GUT scale.  Still,
such large values are possible in a more general model context.  
 It is interesting that the vacuum polarization  measurement 
can be sensitive to fermions with masses well above 1~TeV.

\subsection{Extended Higgs sector}

The models discussed in the previous section introduce the
minimal structure needed to account for the $\Phi$ and the effective
Lagrangian \leqn{effL}.  In this section, we will discuss models that
extend these ideas by postulating a specific relation of the $\Phi$
and the 125~GeV Higgs boson within an extended Higgs sector of fields.

The simplest scheme involves adding a new singlet Higgs boson to the
Standard Model Higgs doublet.  This possibility is explored, for
example, in~\cite{Buttazzo:2015txu,Falkowski:2015swt,Craig:2015lra,Altmannshofer:2015xfo,Cheung:2015cug,Gopalakrishna:2016tku,Benbrik:2015fyz}.   Heavy
vectorlike fermions, as described in the previous section, are needed
to provide the effective Lagrangian couplings \leqn{effL}. 

An $SU(2)$ singlet scalar  boson has the same quantum numbers, after
symmetry breaking, as the 125~GeV Higgs boson and, in general, cannot avoid
mixing with the Higgs boson.  Such mixing effects could appear in
precision measurements of the Higgs boson properties.
 The mixing angle is limited by
electroweak precision measurements to satisfy
   $  \sin\theta < 0.35$
for a singlet mass of 750~GeV~\cite{Gupta:2012mi}.  A stronger limit,
$\sin\theta < 1\%$, comes from the fact that the $\Phi$ can decay to $ZZ$ through mixing
with the Higgs, and this decay is bounded by
observation~\cite{Franceschini:2016gxv}.
The effect of this mixing on the $hWW$ and $hZZ$ couplings is at the
$10^{-4}$ level, and thus probably unobservable.  However, a mixing
angle of this size would still produce percent-level effects in the
$gg$ and $\gamma\gamma$ Higgs couplings.  Also, if $\Phi$
couples to Standard Model fermions such as $b\bar b$ or
$\tau^+\tau^-$,  those couplings of $h$ could be shifted from their
Standard Model values at the percent level.  All of these effects would be
tested in the ILC's
comprehensive program of Higgs boson coupling determinations. 

An interesting possibility raised by \cite{Gopalakrishna:2016tku} is
that a neutral heavy vectorlike fermion could be the particle of dark
matter.
The $\Phi$, if it is a scalar, not a pseudoscalar, would provide an
s-channel resonance  in fermion pair annihilation that would allow the
fermions to have the correct thermal relic density for masses in the
range $300 < m_\psi < 450$~GeV.   Such particles would be extremely
difficult to discover at the LHC, but they would be seen in the
reaction  $\ee\to \gamma \psi \bar\psi$ at the ILC operating an
upgraded energy of 1~TeV~\cite{ILCDM}.

Models of the $\Phi$ as the heavy $H,A$ states of a 2-Higgs-doublet
model
have been presented 
in \cite{Djouadi:2016eyy,Gupta:2015zzs,Altmannshofer:2015xfo,Angelescu:2015uiz,Badziak:2015zez,Becirevic:2015fmu,Bizot:2015qqo}.  Again,
the natural couplings of these particles to $\gamma\gamma$ and $gg$
through Standard Model loops are too small to account for the
observation, so new vectorlike fermions are also needed.  Models of
this type
are constrained at large $\tan\beta$ by the non-observation at the LHC of $\Phi$ decays to
$\tau^+\tau^-$  and at small $\tan\beta$ by the non-observation of
decays to $t\bar t$.   These constraints lead to a preference
for intermediate values of $\tan\beta$, close to $\tan\beta \sim 7 $.
This places the models in the ``wedge'' region of the 2-Higgs-doublet
 model in which it is very difficult for the LHC to observe the heavy
Higgs bosons beyond masses of 500~GeV through more conventional processes
such as $b\bar b \to H,A \to \tau^+\tau^-$.   The enhanced $gg$
coupling of the $\Phi$ gives a  new production mechanism, but still it
might not be seen to decay to heavy fermions.

On the other hand, the 2-Higgs doublet model requires mixing of the
heavy Higgs bosons with the 125~GeV Higgs boson, producing 
shifts in the $\tau\tau$ and $b\bar b$ couplings of the known Higgs
boson.  
Observation of these effects of heavy $H,A$ bosons 
of mass 750~GeV are expected~\cite{Kakizaki:2015zva} to be well within the 5 $\sigma$ reach of
the full ILC program described in \cite{Case}. 

An interesting possibility in the 2-Higgs-doublet model is that the
$H$ and $A$ could be close in mass, so that the $\Phi$ resonance is
actually a double resonance, one scalar and one pseudoscalar
particle.   If the mass difference of these particles is of the order
of tens of GeV, this might explain the broad width required in the
best fit to the ATLAS data.  This scenario can also be realized in
models with $SU(2)$ singlets only~\cite{Wang:2015omi}.  It is possible that
the two separate resonances could be resolved with higher-statistics
$\gamma\gamma$ observations at the LHC.  However, the crucial test of
this model would come in the photon collider experiments described
below in
Section~4.4.   Using transversely polarized photon beams, the shape of
the resonance would shift as the relative beam polarizations were
switched from parallel to orthogonal orientations.

The $\Phi$ has also been interpreted as one of the heavy Higgs bosons
in the NMSSM extended supersymmetric 
model~\cite{Wang:2015omi,Domingo:2016unq,Badziak:2016cfd,Ellwanger:2016qax}.
Here, it is possible to choose parameters such that the LHC production is through $b\bar b$
annihilation and the Higgsinos play the role of the heavy vectorlike
fermions in generating a coupling to $\gamma\gamma$. Generally, for these
parameters,  the mass of the
Higgsino is typically close to $m_\Phi/2$, making Higgsino pairs
discoverable at a 1~TeV energy upgrade of the ILC; however,  for some
parameter sets, the Higgsinos can be as light as 150~GeV.  

One property of
NMSSM models, emphasized in \cite{Domingo:2016unq,Ellwanger:2016qax},
is that the decay of the $\Phi$  can be to very light Higgs states
$a$, with mass close to the $\pi^0$ mass, each of which then decays to
$\gamma\gamma$.
The apparent $\Phi\to \gamma\gamma$ decay would then actually be a
decay to two $\gamma$ pairs of very low mass.  It is unavoidable that
the 125~GeV Higgs boson also has the decay $h\to a a$ at some level,
and the current limit on the branching ratio of the Higgs boson to
this mode, about 1\%,  is already a constraint on the
models.  The High-Luminosity LHC is expected to be sensitive to a 
branching ratio as small as $5\times 10^{-5}$~\cite{Curtin:2013fra}. 
ILC can add to the information on $h\to aa$, since it will be
sensitive to decays of the $a$ to hadrons that might be hidden at LHC.

Two Higgs doublet models have also been introduced to explain a
different aspect of the LHC data, the  suggestion by CMS of a  decay
$h\to \tau \mu$~\cite{Khachatryan:2015kon}.  The second Higgs doublet
has no couplings to fermions except through small flavor mixing
effects. 
 In~\cite{Bizot:2015qqo},
the heavy Higgs scalar in this model is interpreted as the $\Phi$ and appropriate
heavy vectorlike fermions are added to  produce the effective
Lagrangian couplings \leqn{effL}.   It is obvious that, if the CMS
suggestion is confirmed, measurement of the detailed structure of
Higgs flavor violation---both in the quark and lepton sectors---will
be a major task for the ILC.  In addition, mixing between the two
doublets will shift the absolute normalizations of the flavor-diagonal
Higgs couplings, an effect that ILC can probe below the 1\% level. 

\subsection{Bound state of new weakly-coupled constituents}

It has been appreciated for a long time that the top squark might first
be identified through the decay of stoponium to 
$\gamma\gamma$~\cite{Drees:1993yr,Martin:2008sv,Batell:2015zla}.
Pursuing these ideas,
several authors have interpreted the $\Phi$ as a nonrelativistic bound
state of new fermions or
bosons~\cite{Luo:2015yio,Han:2016pab,Kats:2016kuz,Iwamoto:2016ral,Kamenik:2016izk,Foot:2016llc,Ko:2016sht}. Stoponium itself
has a signal cross section about 10 times too small, but this can be
overcome by assuming larger electric charge (e.g., $5/3$) for the
constituents.

 In these models,
the continuum production of the new particles should be discovered at
the LHC.  The effects in precision measurements are small, precisely
because these models are constructed to isolate the $\Phi$ from other 
physics at the TeV scale. At the very least,  the vacuum polarization
corrections from the new particles will produce shifts of order 1\% in
the cross sections for $\ee\to \mu^+\mu^-$ and $\ee\to \ee$, as we
have discussed in a more general context at the end of Section~3.1.
If the new heavy particles have direct couplings to $t$ or $b$, we will see effects
on the Higgs boson and top quark similar to those described above for
heavy vectorlike fermions. 

\subsection{Pion of a new strong interaction sector}

A very different picture of the  origin of a new scalar or
pseudoscalar 
comes from 
models with new strong interactions at multi-TeV energies.  The new
strong interactions might involve a sector of heavy fermions
charged under a confining non-Abelian gauge group.  Such theories
quite naturally contain 
spontaneously broken global chiral symmetries. These lead
to (pseudo-) Nambu-Goldstone bosons (pNGB, generically also called
pions), that are naturally light by the Goldstone mechanism. The
$\Phi$ could be one of these pions.  It is useful to think of it as
analogous to the $\eta$ or
$\eta'$ of QCD.      In many of these models,
 additional  pNGBs are expected to have masses with the reach of the
 LHC, giving the potential to
   further constrain the model space. If the $\Phi$ comes from a decay of another resonance
   with mass larger or equal than 1500 GeV~\cite{Kim:2015ron,Kong,Knapen:2015dap},
   we might already have evidence for one of these states. 

The fermions of the new gauge group
must be charged under the QCD and electroweak gauge groups in order
that the $\Phi$ will have the effective Lagrangian couplings
\leqn{effL}.  If this
is so, the model generates these couplings in the
same way that QCD generates an $\eta \to \gamma\gamma$ coupling~\cite{Nakai,Redi,Harigaya:2016pnu,Harigaya:2016eol}. 
The  precise prediction for the rates and decay patterns of these
fermions depend on the specific model, particularly the confining
gauge group (usually assumed to be some $SU(N)$), the fermion content
and its quantum numbers. Some classification has been done, using
assumptions on minimal flavor violation, the absence of Landau poles
in the gauge couplings as well as the existence of maximally two
diphoton candidates in the models. Still, the model space is
vast. Generically, all kinds of these models, which could be
composite Higgs models, models of partial compositeness, Little Higgs
models or even Twin Higgs models, are all having particular patterns
of deviations in the Higgs (and also top) couplings. Their
measurements -- using the information on the diphoton resonance
gathered at the LHC -- will be an indispensable tool in determining
the specific underlying model.

There is an interesting connection with dark matter.  In models
of dark matter based on models with new strong interactions, the dark
matter particle is typically, the lightest pNGB.   The width of the
$\Phi$ can be enhanced substantially by invisible decays to pairs of
dark matter pNGBs.   If indeed this effect enhances the width of the
$\Phi$, giving better agreement with the current observations, it
argues that the dark matter particle lies below half the $\Phi$ mass
and potentially in the energy range of the ILC.  For an isolated
$\Phi$ resonance, the thermal relic abundance of dark matter is
correct for a dark matter mass of about 300~GeV.   If there are 
two pNGBs close in mass, so that the relic abundance is set by
coannihilation, dark matter masses lower than 100~GeV are favored, and
the dark matter pair threshold would be within the range of the 500
GeV ILC~\cite{Redi}.

\subsection{ Radion of Randall-Sundrum models}

 \def\invfb{ \mbox{fb}^{-1} }

A more concrete model representing effects of new strong interactions
at the TeV energy scale is the Randall-Sundrum (RS) model
\cite{Randall:1999ee}.  In this model, our
 4-dimensional space is extended to  a 5-dimensional warped space with
 4-dimensional boundaries.   Due to the warping of space, excitations
 on the
 boundaries 
naturally have different energy scales.
These boundaries  are called, accordingly, the IR and UV
 branes.  The IR scale is typically about 1 TeV.   
Fields in the 5-dimensional interior give rise to a spectrum  of
 quantum excitations, called Kaluza-Klein (KK) excitations.   The
 spectrum of these KK states extends to infinity, forming the
 so-called ``Kaluza-Klein tower''; however, the lowest-mass states
 typically give the dominant contributions to observable effects.
  The RS model can be considered a dual description of a
  strong-interaction theory above 1~TeV, with the KK excitations
  modelling the strong interaction bound state spectrum.   Or, the RS
  model can be thought of as a 5-dimensional theory of electroweak
  symmetry breaking in its own right. 

The RS model contains a special new scalar particle called the radion.
The radion is a quantum excitation of the position of the IR brane in the
5th dimension.  In models such as 
those of~\cite{Bardhan:2015hcr,Ahmed:2015uqt,Davoudiasl:2015cuo}, the
mass 
of this radion can be of the
order of 500\,GeV, making it reasonable to associate this particle
with the $\Phi$. The phenomenology of the radion depends on the arrangement
of Standard Model and new fields in the 5-dimensional space. The
placement of these fields in the interior or on the boundary is a
matter of model-building, but it is also constrained by electroweak
precision data and the lack of observation, so far, of Standard Model
KK excitations at the LHC.   A typical property of the radion is that
it has a large coupling to $t\bar t$, leading to a large $\Phi\to
t\bar t$ event rate.  The models just cited find parameter sets that
avoid this problem.

There is another type of new particle also predicted in the RS model.
KK excitations of gravity in the 5-dimensional interior can have
couplings of electroweak strength to some Standard Model particles.
It is possible to build models in which the lightest new particle in
an RS model is a spin~2 KK excitation of the graviton.   This gives a
different theory of the $\Phi$ that we will discuss in the next section.

Three effects on ILC precision measurements are expected if
there is an RS extension of the Standard Model that gives the $\Phi$
as a radion.   The first of these effects is Higgs-radion mixing.
The radion is an $SU(2)$-singlet state, and so the phenomenology of
this mixing is similar to that discussed for a singlet Higgs boson in
Section 3.2.  Often, the mixing angle is
very small; for example, \cite{Ahmed:2015uqt} estimates  $\theta_m\sim
m_h^2/6m_\Phi^2 \sim 0.5\%$.    This removes the possibility of
observing shifts in the overall level of Higgs couplings, which are
proportional to 
$\cos\theta_m$, but large couplings of the radion to $gg$ and
$\gamma\gamma$, which give $\sin\theta_m$ shifts of these particular
Higgs couplings, should still be observable. 

The second effect is that of corrections from the tower of KK
states to  the
Higgs boson couplings to heavy species $W$, $Z$ and $t$.   The masses
of the lowest KK states are model-dependent, ranging from a few TeV to
tens of TeV.   However, even at the heavier values, the radiative
corrections to the Higgs couplings can be substantial.  For example,
the models of~\cite{Malm:2014gha} predict shits of the Higgs boson
couplings to vector bosons
$c_{V} = g(hVV)/g(hVV)|_{SM}$ of
the form
\beq
c_W \approx c_Z \approx 1- 0.078\left( {5\ \mbox{TeV}\over  M_{g(1)} } \right)^2.  
\eeq{Malmc}
where $M_{g(1)}$ is the mass of the first Kaluza-Klein excitation of
the gluon. For a 5\,TeV Kaluza-Klein excitation this leads to an
$\sim$8\% deviation.  With the formula \leqn{Malmc}, even a 20~TeV KK
gluon would produce observable effects, for both $W$ and $Z$, in the
ILC precision measurements.
%The model discussed in~\cite{Davoudiasl:2015cuo} includes a 3~TeV 
%Kaluza-Klein gluon, which would lead to an even more pronounced
%effect. 
% (because of details of the Davoudiasl-Zhang construction, this
% statement is not correct)

\begin{figure}
 \begin{center}
    \includegraphics[width=0.99\textwidth]{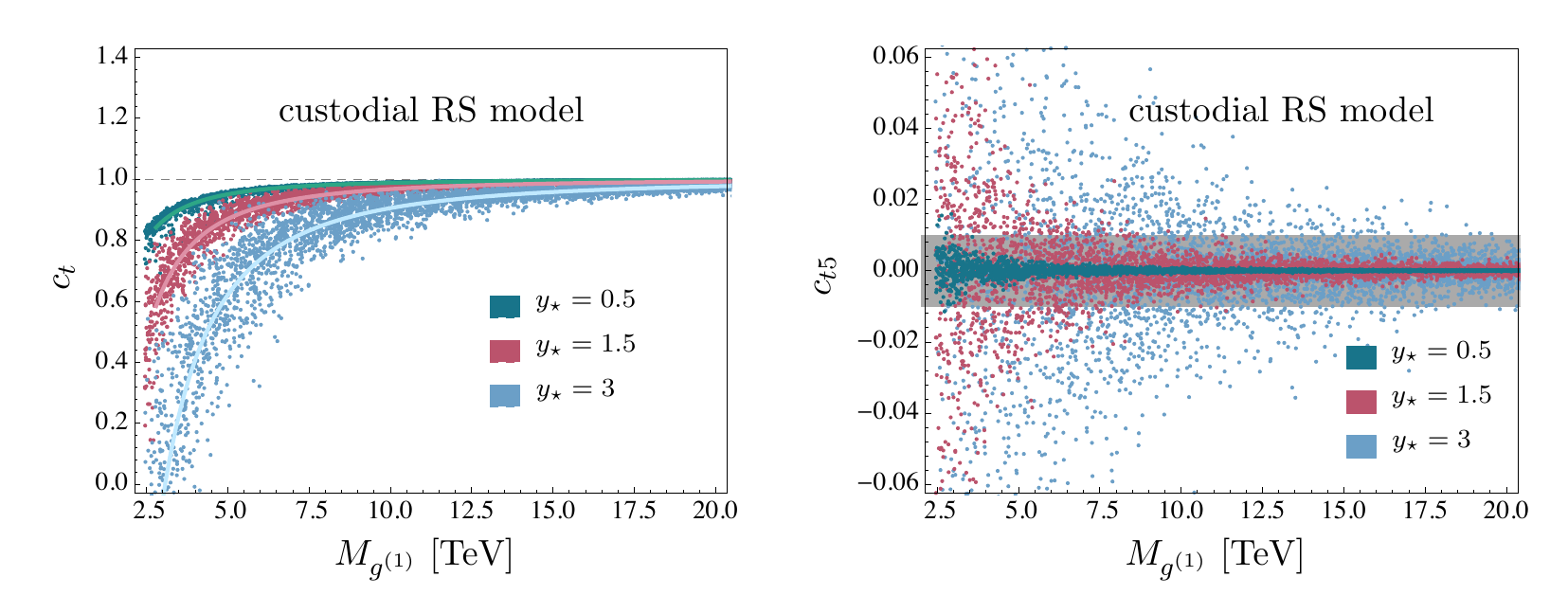}
    \caption{\label{fig:ctct5RSC}Predictions of $CP$-even (left) 
and $CP$-odd (right) Higgs couplings to 
the $t$ quark in a Randall-Sundrum model with a custodial
symmetry. The parameters $c_t$, $c_{t5}$ are defined in
\leqn{explainct}.  The 
point clouds are a scan of the space of Yukawa couplings for three 
values of the free parameter $y_{\ast}$. The gray band on the right hand 
side shows the experimental bound at 90\% on the CP-odd predictions 
derived from the electron dipole moment (EDM). Figure and parts of the 
caption are taken from~\cite{Malm:2014gha}.  }
\end{center}
\end{figure}

The top quark Yukawa coupling should also receive large radiative
corrections from the KK towers of states. Figure~\ref{fig:ctct5RSC}
shows the shifts  of the Higgs boson couplings 
computed in \cite{Malm:2014gha} for
a variety of parameters sets, as a function of the mass of the
lightest KK gluon.  Note that these shifts are in general complex,
leading to CP violation in the $ht\bar t$
coupling at levels observable at the ILC~\cite{Godbole:2011hw}.   
The figure shows the shifts as parameters $c_t$, $c_{t5}$
corresponding to the effective coupling
\beq
    \delta \L =  - {m_t\over v} \ h \  \bigl[  c_t\ \bar t t + i
    c_{t5}\  \bar t \gamma^5 t \bigr]  .
\eeq{explainct}

The third effect is a modification of the electroweak couplings of the
top quark.  In RS models, the top quark wavefunction extends to the IR
side of the 5th dimension, and so the top quark shares some of the
compositeness of the Higgs boson.  In particular, while the
wavefunctions of light fermions peak close to the UV brane, the wavefunction of
the $t_R$ often peaks close to the IR brane.   This may constitute an
elegant explanation of the striking mass hierarchy in the fermion
sector.   A consequence of this effect is that the couplings of the
top quark to the $Z$ boson are expected to have large shifts due to
mixing with KK states, of independent size for $t_L$ and $t_R$.
Models of this effect are described, for example, in
\cite{Djouadi:2006rk,Carena:2006bn}. 

One of the well-appreciated features of the ILC is that the $t\bar tZ$
couplings enter the expressions for the  production of top quark pairs
in $\ee\to t\bar t$.  The couplings to $t_L$ and $t_R$ can be
separately measured using polarized beams~\cite{Amjad:2015mma}.   In
\cite{Case}, these measurements are reviewed, and errors of order  1\%
on the separate $Z$ couplings to $t_L$ and $t_R$ are estimated for
the full ILC program.   As discussed there, these accuracies suffice
not only to observe the effect but also to provide significant
discrimination among models.

\subsection{Graviton of Randall-Sundrum models}

The $\Phi$  resonance could be a
particle of spin 2 rather than a particle of spin 0. 
At the LHC,  the spin of the resonance can be determined 
by studying the angular distribution of the photons.  
Demonstrating that the photon distribution in the rest frame 
of the diphoton system is not isotropic, but instead is strongly peaked
 in the beam axis direction,  would be a smoking gun for the spin-2 nature of the resonance.  
 
One example of a theory that provides a spin-2 resonance with
properties of the $\Phi$
 is the Randall-Sundrum (RS) model~\cite{Randall:1999ee}, in the
 region of parameter space where the lightest  KK graviton excitation
 is the lightest new particle.  By the introduction of additional
 interactions on the 4-dimensional  boundaries of the warped space,
 it is possible to realize parameters in which the KK graviton is
lighter than the RS radion excitation described in the previous
subsection.    In such a construction, there is a danger that the
radion may become a negative-metric (ghost) particle; however, the papers
cited below find parameter sets that avoid this problem. 

A key issue relevant to the interpretation of the $\Phi$ as an RS
graviton is the question of whether the resonance couples to $\ee$
pairs.   In the simplest scenario, the  entire Standard Model  is localized on the
 IR brane \cite{Giddings:2016sfr}. In this case, the $\Phi$ would
 couple to the energy-momentum tensor of the Standard Model and thus
 would couple with the same strength to $\ee$ and $\mu^+\mu^-$ as to
 $\gamma\gamma$.
Then the $\Phi$ would be expected to appear as a resonance in the
Drell-Yan process.   This is not observed and, while this model is not
yet excluded,  it provides a nontrivial constraint.
However, it is also possible to  have non-universal couplings of the
spin-2 resonance to the SM if one allows the SM fields to also
propagate in 5 dimensions.   In a construction of this type, lighter
quarks and leptons have 5-dimensional wavefunction localized further
from the IR brane.   This leads to smaller Yukawa couplings and
smaller fermion masses; thus, the construction can be used to address
 the fermion mass puzzle.    Wavefunctions further from the IR brane
 also couple less strongly to the lowest KK graviton excitations, thus
 relaxing the constraint from the Drell-Yan
 process~\cite{Falkowski:2016glr,Hewett:2016omf,Carmona:2016jhr,Dillon:2016fgw}.  

The RS graviton couples to photons and gluons, and the parameters 
of the model can be adjusted to correctly reproduce the LHC observations of the diphoton resonance. 
Other decay modes should also be observed.
In particular, if the RS framework  plays a role in addressing the 
hierarchy problem, also $WW$, $ZZ$, $hh$, and $t \bar t$ decays 
are expected to occur with a significant branching fraction.
Table~\ref{tab:bench}
  shows the KK graviton branching fractions in
several benchmark models, including the IR model  from
\cite{Giddings:2016sfr} and four models with  SM fields in the bulk
defined in  \cite{Falkowski:2016glr}.

\begin{table}[h]
\begin{center}
\begin{tabular}{|c|ccccc|}
\hline%$f$ & Br$(X \to f)$ [\%]  &  ${{\rm Br}(X \to f) \over {\rm Br}(X \to \gamma \gamma)}$ 
& \ IR \ & MIN & MED & MAX  & GMAX
\\  \hline\hline
${\rm Br}(X \to \gamma \gamma)$ [\%]  & 4.3 &  8.5 &  7.0  & 0.5 & 2.3 
\\  %\hline
${\rm Br}(X \to  ZZ )$  [\%]  & 4.8 & 7.9 &  7.8 &  2.9   & 12 
\\  %\hline
${\rm Br}(X \to WW)$  [\%]  &9.5 &  16 &  15   & 5.6 & 21  
\\  %\hline
${\rm Br}(X \to Z\gamma)$  [\%]  & 0 & 0 &  0 &  0 & 1.1 
\\  %\hline
${\rm Br}(X \to hh)$   [\%]  & 0.3 & 0 &  0.4   & 1.4 &6.9 
\\  %\hline
${\rm Br}(X \to t\bar t)$  [\%]   &5.1 & 0 &  8.3  & 85 & 56 
\\  %\hline
${\rm Br}(X \to b\bar b)$  [\%]   &6.4  & 0 &  5.2      & 0.4  & 0.04 
\\  %\hline
${\rm Br}(X \to jj ) $  [\%]   &66 & 68 & 61    & 4.5  & 0.5 
\\  %\hline
${\rm Br}(X \to e^+e^-)$  [\%]   & 2.1 & 0 &  0  & 0 & 0 
\\  \hline  \hline  
$\Gamma(X \to \gamma \gamma)$[MeV]  & 0.25  &0.15 & 0.18 & 2.5 & 25
\\  
$\Gamma(X \to {\rm tot})$[MeV]  & 5.7  &1.8  & 2.6 & 500  & 1060 
\\  
PLC:   $\sigma_{eff}(\gamma \gamma \to X)$ [fb]  & 40  &24  & 29  & 400  & 4000
\\  
LC:  $\sigma(e^+ e^- \to X)$~[pb] & 0.4   & 0  & 0 & 0  & 0
\\  \hline   
\end{tabular}
\caption{
\label{tab:bench}
Observables  for the IR model defined in Ref.~\cite{Giddings:2016sfr}
and for the MIN, MED, MAX, and GMAX benchmarks in the bulk RS scenario
  defined in  Ref.~\cite{Falkowski:2016glr}. }
\end{center}
\end{table}

From this discussion, we see that the size of the coupling of the
$\Phi$ to $\ee$ is a key test that discriminates RS models on the basis
of the localization of wavefunctions in
the 5th dimension.  We have pointed out in Section 2 that the $\Phi$
coupling to $\ee$ is constrained by the non-observation of the $\Phi$
in the Drell-Yan spectrum, but it could still be comparable to the
$\gamma\gamma$ coupling.   Thus, the possibility
is open that the $\Phi$ could be observed as a resonance in $\ee$
collisions. 

One might hope that, in $\ee$ collisions at 500~GeV, the $\Phi$ might
be observable as a spin-2 contact interaction in
$\ee\to \gamma\gamma$ or $b\bar b$.   Unfortunately, for the models in
Table~\ref{tab:bench}, the effects on the angular distributions in the
2-body final state are at most of the order of $10^{-4}$.   However,
the presence of 
an RS model at 1~TeV will be recognizable at the ILC, since it will
produce the spectrum of effects on precision measurements described in
the previous section.

With an energy upgrade of the ILC to 1~TeV, it will be possible to
explore
for the $\Phi$ resonance in $\ee$ annihilation more directly.   We
will discuss the phenomenology of a spin-2 $\Phi$ resonance in Section
4.5.

\subsection{Summary}

The discussion of this section is summarized in
Table~\ref{tab:anomalies}.   For each of the models that we have
discussed in this section, we mark in this table the precision  Higgs,
top or $\ee$ measurement available from the ILC in which a significant
deviation from the Standard Model would be expected. The
observation of these anomalies would put us on the right path toward
building concrete theories to explain the $\Phi$.  The values of the
anomalies will fix explicit parameters in these theories.  More
information would be available if the anomalies observed at the ILC
could be correlated with the properties of new particles discovered at
the LHC.  We hope for such discoveries, but the discovery of further new
particles at the LHC is not guaranteed in any scenario.  The table
makes clear that, independently of any further information from the
LHC, precision measurements at the ILC will give many new pieces of
information on the origin and nature of the $\Phi$ resonance.

\begin{table}
\begin{center}
\begin{tabular}{l|l|c|c|c|c|c|c|c|c|c} 
Sect. & & $ hWW$ & $h b\bar b$ & $h\gamma\gamma$ & $ht\bar t$ & $h\to$ &$ h\tau\mu$& $t\bar t Z$ 
 &$ee\to$ & $ee \to $  \\
&& $ h ZZ$ & $h\tau\tau$ &$hgg$ &  &\ invis.  & & & $ee,\mu\mu$ & $\gamma +$ invis. \\ \hline
3.1 & Vectorlike &    & & & & & && \\ 
&\ \ fermions & &  X&X&X&& &X&X& \\  \hline
3.2 & Higgs      &    &&&&&&&& \\
&\ \ singlet & &X&X &X &&&X&\\   \hline
3.2 & 2 Higgs    &    &&&&& &&&\\ 
&\ \ doublet & X &X&X&X&&&&&\\   \hline
3.2 &NMSSM                &  &&&&&&&\\ 
 &              & X  &X&X&X&X&&&&X\\ \hline
3.2 & Flavored   &    &&&&&& &\\ 
& \ \ Higgs   &X &  X &X& &&X&&& \\  \hline
3.3&Bound &&   && &&&& &\\
&\ \ state &   && & & &&&X& \\ \hline
3.4 &Pion of                    &   &&&&&&&&\\ 
&new forces                   &   &X&X&X&X&&X&X&X \\ \hline
3.5 &RS            &  &&&&& &&\\  
&\ \ radion              & X& X &X&X&&&X && \\ \hline
3.6 &RS        & &&&&&&& \\ 
&\ \ graviton                   & X &X &&X&&& X&& \\ \hline
\end{tabular}
\caption{\label{tab:anomalies}  Anomalies in precision measurements expected to be visible at the
  ILC for the models of the $\Phi$ discussed in Section 3 of this report.}
\end{center} 
\end{table}

\section{Observation of the $\Phi$ in
 $\gmgm$ and $\ee $ collisions}

Up to this point, we have been discussing only tests of theories of
the $\Phi$ available at energies of 500~GeV and below.    However, the
ILC TDR also envisioned an energy upgrade to 1~TeV~\cite{ILCTDR}.   
  This upgrade
would give the possibility of producing the $\Phi$ directly. 
 Since the $\Phi$ is observed
in the decay $\Phi\to \gamma\gamma$, at least one of the couplings 
$A_1$ or $A_2$ in \leqn{effL} must be nonzero.  Then there are nonzero
cross sections for production of the $\Phi$ both 
in $e^+e^-$ and in $\gamma\gamma$ collisions.   If the $\Phi$
resonance can be observed, these processes will give  access to the
full range of decay modes of the resonance, just as the ILC is
expected to allow the study of the full set of decay modes of the
125~GeV Higgs boson.

Four distinct processes can be studied.   First $\Phi$ can be
produced from $\ee$ beams
via the associated production with $\gamma$ or $Z$,
\beq
e^+e^- \to \Phi V   \  , \quad 
\mbox{with} \quad V=\gamma , Z \   . 
\eeqn
These processes can be  characterized by the observation of a 
 monochromatic electroweak gauge boson.

 Second, $\Phi$ can be produced 
via the vector boson fusion processes~\cite{Ito:2016zkz}
\beq
    e^+e^- \to \Phi e^+e^-\   , \quad \Phi \bar{\nu}_e \nu_e \ .
\eeqn
These processes can be identified by the characteristic transverse
momentum of order $m_W$ imparted to  the $\Phi$.   The reaction 
$e^+e^-\rightarrow\Phi e^+e^-$ is also likely to have energetic $e^\pm$
observed at small scattering angles.

Third, the ILC at 1~TeV can be used as the basis of a 
photon-photon linear collider (PLC).  In this facility, 
the $\Phi$ would be  produced in $\gamma\gamma$
collision \cite{Ito:2016zkz,Djouadi:2016eyy,ZGHe,Richard:2016nhm}
\beq
\gamma\gamma \rightarrow \Phi \ . 
\eeqn

Finally, the possibility of a direct coupling of $\Phi$ to $\ee$
raises the possiblity of observing the resonant production $\ee\to
\Phi$.  

In this section, we will discuss the cross sections for $\Phi$
production and the possibility of $\Phi$ observation in these
processes.   The cross sections that we compute will be based on the
effective Lagrangian \leqn{effL} and will be proportional to
$\Gamma(\Phi\to \gamma\gamma)$.  For numerical estimates, we will use
the conservative reference value
$\Gamma(\Phi\to \gamma\gamma) = 0.5$~MeV, as discussed below
\leqn{refGamma}.  We have pointed out in Section 2 and in
Fig.~\ref{fig:CERN}  that the value of $\Gamma(\Phi\to \gamma\gamma)$
could
potentially be three orders of magnitude larger than this value.  In
that case, the cross sections computed  in this paper would be larger
by the same factor.   The reader should keep this in mind in 
evaluating 
the rate estimates that  we present here. 

\subsection{$e^+e^- \rightarrow \Phi \gamma$ or $\Phi Z$}

From the effective Lagrangian \leqn{effL}, it is straightforward to
work out the cross section for $e^+e^- \rightarrow \Phi V$.  
These cross sections simplify dramatically in the limit $s/\mz^2 \gg
1$, which actually applies for $\sqrt{s} \geq  750$ GeV.   In that limit,
the polarized cross sections are
\beq
  {d\sigma\over d\cos\theta} =  {\alpha_w^3 \over  32}  | C|^2
  (1+ \cos^2\theta)  (1 - m_\Phi^2/s)^3 \ , 
\eeq{csPhiforms}
where  the coefficients $C$ are given by 
\beqa
   C(e^-_Le^+_R \to \Phi \gamma)  &=&  \biggl( {s_w\over 2} A_2  +
   {s_w^3\over 2 c_w^2} A_1 \biggr) \CR
   C(e^-_Re^+_L \to \Phi \gamma)  &=&  \biggl( 
   {s_w^3\over c_w^2} A_1 \biggr) \CR
   C(e^-_Le^+_R \to \Phi Z) & = &\biggl({c_w\over 2}A_2  -
   {s_w^4\over 2 c_w^3} A_1 \biggr) \CR
   C(e^-_Re^+_L \to \Phi Z) & =&  \biggl( -
   {s_w^4\over c_w^3} A_1 \biggr) 
\eeqan
For $A_2 \sim (500~\mbox{GeV})^{-2}$ as suggested by \leqn{refGamma}, this
gives cross sections from the $e^-_Le^+_R$ state at $\sqrt{s} = 1$~TeV
of order of tens of  ab, which
is small but promises a finite sample of events.  If higher energies
are accessible, the cross section increases rapidly as $\beta\to 1$.  In this
approximation, the same cross section formulae apply for a scalar or a
pseudoscalar $\Phi$, using for the pseudoscalar case the modified
Lagrangian described in the text below \leqn{effL}.  

\begin{figure}
 \centering
 \includegraphics[width=.95\hsize]{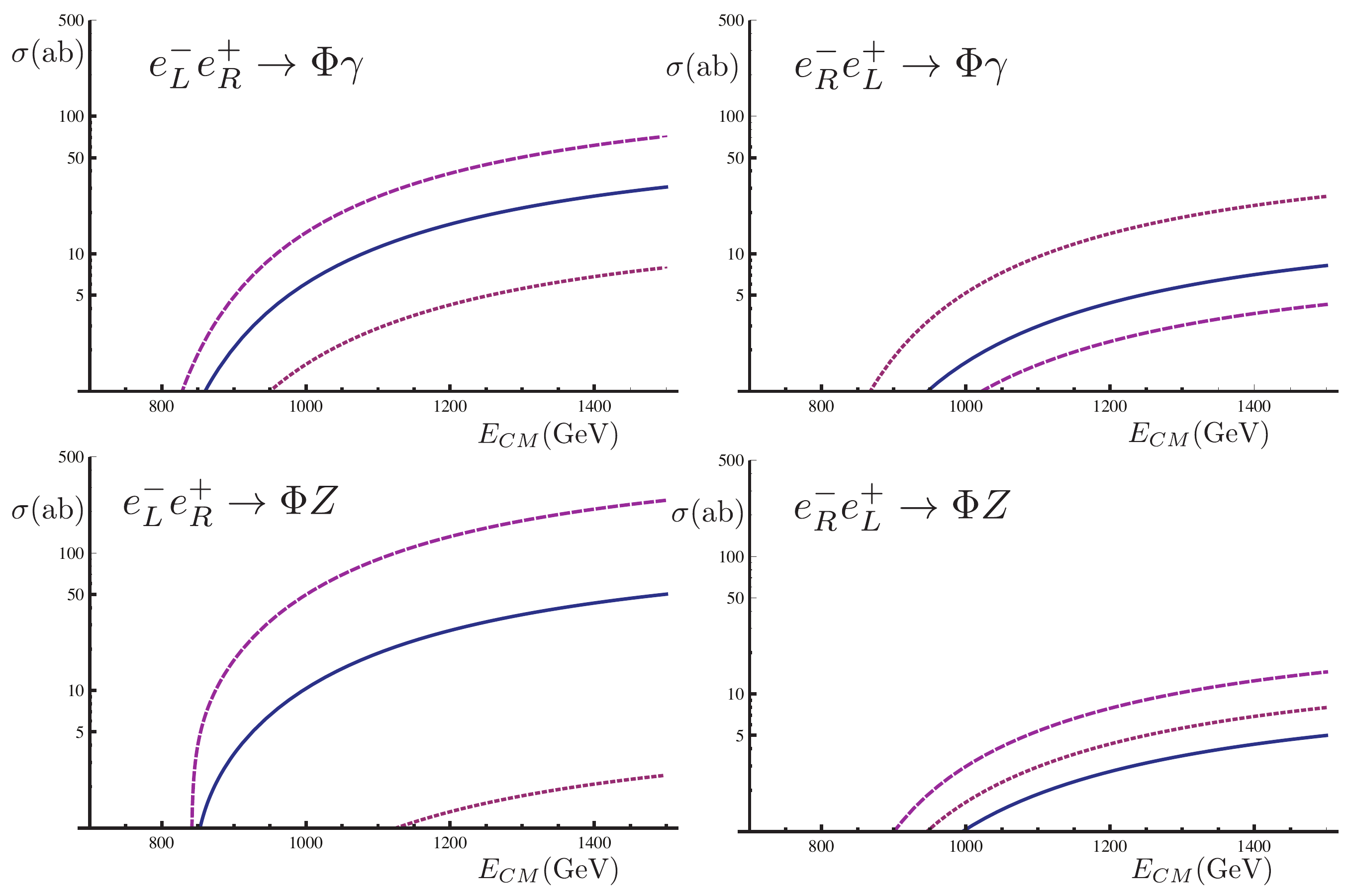}
  \caption{Total cross sections for the process $\ee\to \Phi \gamma$,
    $\ee\to \Phi Z$, for polarized beams, in ab, as a function of
 the collision energy.  The different curves refer to different values
 of $R$ in \leqn{Rdef}:  dashed: $R =
 \infty$ (pure $ A_2$; solid: $R
 = 1$; dotted: $R = 0$ (pure $A_1$). In each case, the cross sections
 are computed assuming 80\% electron beam polarization and 30\%
 positron beam polarization. ISR and beamstrahlung are not
 included. These cross sections are based on a conservative value of
 $\Gamma(\Phi\to \gamma\gamma)$, as discussed in the text. }
  \label{fig:xsec_ax}
\end{figure}

The cross sections of processes $\ee\to \Phi \gamma$ and $\ee\to \Phi Z$ for different
beam polarisations are shown
in Fig.~\ref{fig:xsec_ax}.   These figures use the complete formulae~\cite{Ito:2016kvw,FTY}
rather than the approximations \leqn{csPhiforms}. 
The various curves in each figure correspond to different values of
the parameter $R$ given by 
\beq
R= {A_2/ A_1} \ ,
\eeq{Rdef}
assuming the $\Phi\to \gamma\gamma$ partial width given in 
\leqn{refGamma}.  (The results for the upper limit $R = 12$ quoted in
Section 2 are close to those for $R = \infty$.)  The figures include the expected 80\% electron beam
polarization and 30\% positron beam polarization.

The expected cross sections are small in the most conservative
scenario.  However, if a substantial event sample can be gathered,
these processes potentially give a large amount of information about
the $\Phi$.   The most important feature is that 
the $\Phi$ can be tagged by the  $\gamma$ or
$Z$ recoiling against the $\Phi$
without any requirement on the $\Phi$ decay mode.  The situation
is similar to that for the Higgs boson in $\ee\to Zh$.  This allows
us to determine
the absoute partial widths  $\Gamma(\Phi\to \gamma\gamma)$, 
 $\Gamma(\Phi\to \gamma Z)$, and   $\Gamma(\Phi\to ZZ)$, and, from
 these, the total width of the $\Phi$ and the branching ratios to any
 other observed decay channel.   The recoil method is sensitive to
 $\Phi$ decays to invisible particles and to other exotic final states
 that are 
 difficult to reconstruct.

The CP property of the $\Phi$ can be studied in $\ee\to \Phi Z$, with
$Z$ 
decaying hadronically.  This might usefully supplement CP and spin
information obtained from studies of $\Phi\to ZZ\to 4$~leptons at the LHC.

\subsection{$e^+e^- \rightarrow \Phi e^+e^-$}

The reaction $e^+e^- \rightarrow \Phi e^+e^-$ can proceed through
$\gamma\gamma$ or $ZZ$ fusion.   The leading contribution comes from
$\gamma\gamma$ fusion with small scattering angles of the electron and
positron, which can be estimated by using the 
equivalent photon approximation \cite{Budnev:1974de}.  However, the
resulting cross section are very low.   We expect a total rate of
22~ab.   To isolate the signal from background, it is useful to
detect either or both of the $e^-$ and $e^+$ in forward detectors.
If we require that at least one of the scattered particles has
scattering angle larger than 10 mrad and energy larger than 50~GeV, we
find a cross section of 8 ab.   Requiring both the $e^-$ and $e^+$ to
meet this criterion gives a cross section of 1 ab.

\subsection{Interlude: PLC}

If the $\Phi$ decays to $\gamma\gamma$, it must also be produced in
$\gamma\gamma$ collisions. To observe this process with sufficient
rate, the ILC should be converted to a Photon Linear Collider (PLC),
by Compton backscattering of laser beams from the electron beams.    Analyses of the $\Phi$
cross section in $\gamma\gamma$ in this setting  have  been given
 in \cite{Ito:2016zkz,Djouadi:2016eyy,ZGHe,Richard:2016nhm}.  Useful
 background on the accelerator physics of a $\gamma\gamma$ collider can be found
 in the relevant volume of the TESLA TDR~\cite{Badelek:2001xb}.

A useful figure of merit to understand the observability of the $\Phi$
is the  size of $\Gamma(\Phi\to \gamma\gamma)/m_\Phi$, which can be
compared to the similar ratio for the 125~GeV Higgs boson.   The
values are
\beq
  {\Gamma(\Phi\to \gamma\gamma)\over m_\Phi} \geq  7\times 10^{-7} \ , 
  \qquad
 {\Gamma(h\to \gamma\gamma)\over m_h} =  7\times 10^{-8} \ .
\eeq{compare}
Thus, we expect that the $\Phi$ will be easier to observe above
background than the 125~GeV Higgs boson, whose production at a
$\gamma\gamma$ collider has been studied in detail in \cite{Asner}. 

A $\gamma\gamma$ collider would be based on a $e^-e^-$ operation of
the ILC, at an energy about 25\% higher than the energy of the $\Phi$ resonance.
The most commonly discussed operating point for a $\gamma\gamma$
collider is at  the parameter of the Compton scattering process
\beq    x =   {s_{e\gamma}\over m_e^2}  = 15.3  \bigl({ E_{beam}\over
  \mbox{TeV}}\bigr)\bigl({\omega\over \mbox{eV}}\bigr) 
\eeq{xdefin}
having the value   $x = 4.8$~\cite{Telnov}.
For operation with a maximum $\gamma\gamma$ center of mass energy of
800~GeV, this  requires a 1~TeV $e^-e^-$ collider and  a laser of
wavelength $2\mu$m and average power about 100~kW.   The time structure
of the laser must be matched to the time structure of the ILC beams.
A possible solution to this problem is to construct an 
an optical cavity surrounding the ILC detector whose length is close
to the time separation within a train of ILC pulses~\cite{Monig}. 
High-power lasers of 100~kW are available today  at 
$1\mu$ wavelength.  This would require working at value of $x$ that is
acceptable though not optimal for the PLC.  The
technology of high-power lasers is advancing rapidly though, and it is
likely that a laser completely appropriate to the PLC will be
developed by the time the ILC is built.

There is one serious  ILC accelerator issue relevant to the PLC:  To accomodate a
$\gamma\gamma$ collider, the beam crossing angle of the ILC would need
to be increased from the specification of 14~mrad
given in the ILC TDR to incorporate the Compton backscattering system
and  beam dump.  In principle, the crossing angle of the ILC
could be increased to 25 mrad at the time of the energy upgrade by
rebuilding the interaction region.   However, it might be advantageous
to plan for this from the beginning by increasing the initial crossing
angle to 20 mrad, if this can be done without compromising the
hermeticity
in the forward region.   Further study is needed to find the best path.  

\subsection{$\gamma\gamma \rightarrow \Phi$ at the PLC}

For monochromatic photon beams, the cross
section for the process $\gamma\gamma\rightarrow \Phi$ is given by
\beq
  \hat{\sigma}
  (\gamma\gamma\rightarrow \Phi) =
  {16\pi m_\Phi^2  \over s}(2J+1)
  {\Gamma (\Phi \rightarrow \gamma \gamma)\over 
  (s-m_\Phi^2)^2 + m_\Phi^2 \Gamma_\Phi^2} (1 \pm \lambda_1\lambda_2),
\eeqn
where $s$ is the square of the $\gamma\gamma $ center of mass energy and $J$ is the
spin of the resonance. The parameters $\lambda_1$, $\lambda_2$ are the
polarizations of the two photon beams; their relative sign should be
chosen corresponding to the spin of the resonance.
Except
in the most optimistic case, the width of the $\Phi$ resonance would
be small compared to the intrinsic energy spread of
Compton-backscattered photon beams. In \leqn{effLgg} below, we 
define a $\gamma\gamma$ luminosity appropriate to the evaluation of 
Standard Model backgrounds. For use with this luminosity estimate, the
effective resonance cross section is
\beq
    \sigma_{eff} = 4\pi^2 (2J+1) {\Gamma(\Phi\to
      \gamma\gamma)\over m_\Phi^2} (1 \pm \lambda_1\lambda_2)
          \cdot\biggl( {1\over 2E_0 \sqrt{2\pi} \Delta_z }\biggr) \ , 
\eeq{PLCeffsigma}
where $E_0$ is the electron beam energy and $\Delta_z = 2.5\%$ is the
width of the distribution in $z = \sqrt{ s(\gamma\gamma)/s(e^-e^-)}$
in a Gaussian approximation~\cite{Richard:2016nhm}. As we
will discuss below, we estimate the  integrated luminosity sample for
 the PLC to be  about 900~fb$^{-1}$. 

Evaluating \leqn{PLCeffsigma} with a $\gamma\gamma$ width of 0.5~MeV,
we find
\beq
       \sigma_{eff} =   440~\mbox{fb} \ , 
\eeq{firstsigeff}
corresponding to 400,000 $\Phi$ events for the expected data sample. More
precisely,
this is the cross section for $\gamma\gamma\to \Phi\to gg$, set by 
\leqn{refrat}.  If other possible decay modes, such as $ZZ$ or $t\bar
t$, have substantial  branching ratios, the rates to those channels add to
the value in \leqn{firstsigeff}. 

\begin{figure}
 \centering
\includegraphics[width=0.6\hsize]{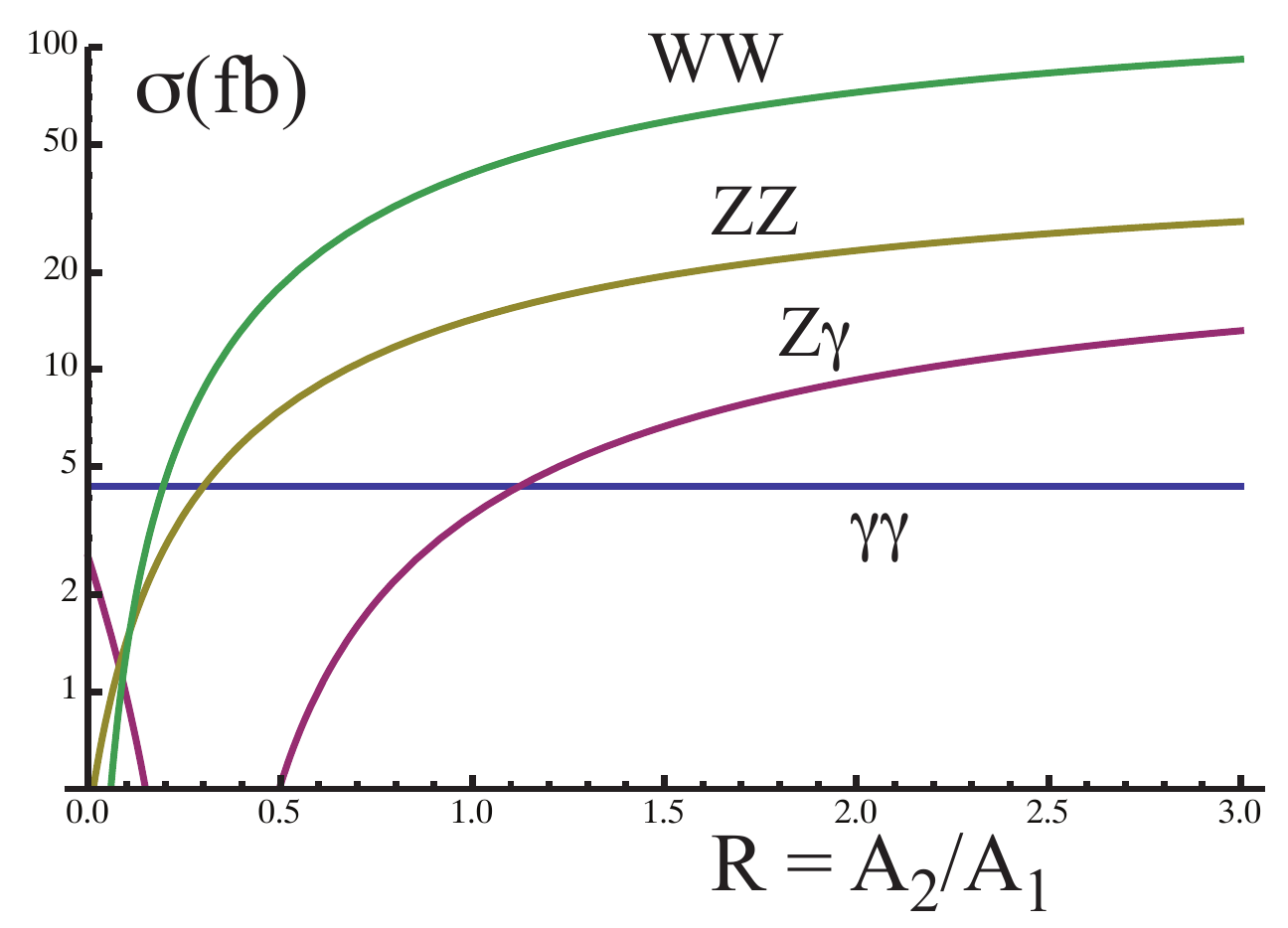}
  \caption{Effective cross sections for $\gamma\gamma\to \Phi \to VV$
    as a function of the ratio of effective Lagrangian couplings
    $R = A_2/A_1$ .  The normalization of these cross sections assumes the
    conservative value of $\Gamma(\Phi\to \gamma\gamma)$, shown in
    Fig.~\ref{fig:CERN}
and an estimate of the total width
    corresponding to  $BR(\Phi\to \gamma\gamma) = 10^{-2}$. }
  \label{fig:gamZW}
\end{figure}

The PLC gives a very clean experimental setting for the measurement of
the relative $\Phi$  decay branching ratios to electroweak gauge
bosons $\gamma\gamma$, $\gamma Z$, $ZZ$, $W^+W^-$.   The cross
sections for these four reactions, based on the conservative estimate
\leqn{Avals} are shown in Fig.~\ref{fig:gamZW} as a function of the
ratio $R = A_2/A_1$.  This is the ideal way to measure this ratio
of effective Lagrangian couplings, which is central to the
interpretation of any model of the $\Phi$.  The number of $\Phi$
events will be an order of magnitude smaller than 
 that expected from the High-Luminosity LHC, but this will be
 compensated by a lower background.  Also, $Z$ and $W$ will be
 detected in their hadronic decay modes, using very similar analyses
 with many systematic errors cancelling in the ratio of rates.

These estimates of signal rates should be compared to estimates of
background rates from pair production in $\gamma\gamma$ reactions.
These rates are evaluated carefully in
\cite{Ito:2016zkz,Richard:2016nhm}; here we will only summarize the
results.  We assume that hadronic background events  are selected to
have the $\Phi$ invariant mass within a detector resolution of 5\%.
Light-quark and gluon pairs cannot be distinguished.  On the other
hand, the very efficient $b$ tagging expected from the ILC detectors
implies that $b$ quarks will make a negligible contribution to the
light jet rates.  We quote the light-quark background rates for 
photon beam polarization $\lambda_1\lambda_2
\approx +1$, which suppresses light-quark pair production.  This is
appropriate to a spin 0 resonance, while for a spin 2 resonance, the
light $q\bar q$
background rates will be about a factor 3 higher. 
 We also impose a cut $|\cos\theta|
< 0.8$ to decrease the background from light fermion pairs and $WW$,
which are strongly forward-peaked.  This angular cut also has the
pleasing effect of narrowing the distribution in $z$, allowing us to
achieve
the width $\Delta_z = 2.5\%$ quoted below \leqn{PLCeffsigma}. 

The luminosity of a $\gamma\gamma$ collider is computed from the $e^-e^-$
luminosity
and the energy spectrum of backscattered Compton 
photons~\cite{Ginzburg:1982yr,Ginzburg:1999wz}.  In $\ee$
operation of the ILC, it is necessary to choose a beam shape for the
colliding beams that compromises between maximizing luminosity and
minimizing beam disruption and beamstrahlung in the collision.   For a
$\gamma\gamma$ collider, the latter restriction is relaxed, and one
can choose a scheme with overall tighter
focussing~\cite{Telnov:1995hc}.   The ``geometrical luminosity'' in
this scheme can be higher than the expected $\ee$ luminosity by a
factor greater than 3. On the other hand, the useful $\gamma\gamma$
luminosity is derated from the geometrical luminosity by a large
factor~\cite{Badelek:2001xb}.
For the background estimates described above, this useful luminosity
is~\cite{Richard:2016nhm}
\beq
       {\cal L}_{\gamma\gamma} = {\cal L}_{geom} \cdot (0.58) \
       \sqrt{2\pi}\Delta_z = {\cal L}_{geom} \cdot  3.6\%  \  .
\eeq{effLgg}
For operation of the ILC at 1 TeV, the expected $\ee$ event 
sample is 8~ab$^{-1}$~\cite{Barklow:2015tja}.   The corresponding $e^-e^-$
geometrical luminosity would be about 25~ab$^{-1}$, giving a
$\gamma\gamma$ integrated luminosity of  900~fb$^{-1}$, as quoted above.

With this understanding, the
rates for a variety of $\gamma\gamma$ 
backgrounds are shown in  Table~\ref{tab:backgrounds}.
The analysis is done for the case of a spin 0 resonance.
The first line of the table gives the expected background cross
section in fb.   The second line gives the value, for the final state $A$,
of the ratio of branching ratios
\beq
    BRR(A) = \Gamma(\Phi\to A)/\Gamma(\Phi \to gg) \ ,
\eeq{BRRdef}
for 
which $S/\sqrt{B} > 5$   (5 \ $\sigma$ observation)  with the 
900~fb$^{-1}$ data set. Note that, if $\Phi\to gg$ is not the dominant
decay mode of the $\Phi$,  $ BRR(A) >  BR(\Phi \to A)$ and the
sensitivity to all final states will be comparably greater.   The
value for  $BRR(gg)$ in the table, 0.3\%,  should be interpreted
as the statement that the expected rate for $\gamma\gamma\to \Phi\to
gg$ is 300 times the level needed for a 5~$\sigma$ observation.

\begin{table}
\begin{center}
\begin{tabular}{l|ccccccccccc}
  & \  $qq+gg$ & $bb$ & $tt$ &$ ee/\mu\mu/\tau\tau$ & $\gamma\gamma $ &
 $  Z\gamma$ & $ZZ$ & $hh$ & $WW$ & $Zh$ \\ \hline
$\sigma_{bkgd}$ (fb) & 46  &  2 &   760 & 40  & 20 & 20 & 20 & $<0.4$& 7600 &
1 \\
$BRR_{5\sigma}$ & 0.3\% & 0.07\% &1.3\% & 0.3\% & 0.2\% &0.2\% &0.2\% &0.03\% &
4\% &0.04\% \\
\end{tabular}
\caption{ \label{tab:backgrounds}   Standard Model background cross
  sections for the observation of decays of a spin 0 resonance $\Phi$ at the
  PLC.  
 The second
  line gives the braching ratio relative to $BR(\Phi\to gg)$ (see
 \leqn{BRRdef}) for a 5
  $\sigma$ observation with a 900~fb$^{-1}$ data set as described in
  the text.}
\end{center}
\end{table}

The LHC will have similar sensitivity for some of these channels.
The preliminary LHC observation is already close to the sensitivity
quoted in the table for $\gamma\gamma$, and the LHC sensitivity already
exceeds the estimate given for $\ee$ and $\mu^+\mu^-$ (with, however,
no observation of the resonance).   On the other hand, the capability for
direct observation of the $gg$ decay and the sensitivity to $b\bar b$,
$t\bar t$, and Higgs modes far exceed what will be possible at the
LHC.

In the case that the $\Phi$ has spin 2, the production cross section
\leqn{PLCeffsigma} is 5 times higher.  However, the signal to
background ratio is also decreased by two effects.   First, a spin 2
particle is produced in $\gamma\gamma$ collisions in the $J^3 = \pm
2$ $\gamma\gamma$ polarization states, which also have a higher  cross
section for light quark pair production.   Second, the signal cross
section is more forward-peaked, so that the $|\cos\theta|
< 0.8$  cut has lower efficiency for the signal  (65\% for $\Phi\to
gg$, assuming an energy-momentum tensor-like couplings, compared to
80\% for the spin 0 case)~\cite{Panico:2016ary}. 
 Taking these factors into account, we
present the observable branching fractions for the spin 2 case
in Table~\ref{tab:backgroundstwo}.   Overall, the prospects of
measurement of many branching fractions of the $\Phi$ are also quite
optimistic in the spin 2 case.

\begin{table}
\begin{center}
\begin{tabular}{l|ccccccccccc}
  & \  $qq+gg$ & $bb$ & $tt$ &$ ee/\mu\mu $&$ \tau\tau$ & $\gamma\gamma $ &
 $  Z\gamma$ & $ZZ$ & $hh$ & $WW$ \\ \hline
$\sigma_{bkgd}$ (fb) & 760  &  11 &  3800 & 110 & 1300  & 20 & 450& 200 & 40&
27000 \\
$BRR_{5\sigma}$ & 0.3\% & 0.03\% &0.5\% & 0.08\% & 0.2\% &0.05\% &0.4\% &0.17\% &
0.04\% &2\% \\
\end{tabular}
\caption{ \label{tab:backgroundstwo}   Standard Model background cross
  sections for the observation of decays of a spin 2 resonance $\Phi$ at the
  PLC.  
The second
  line gives the braching ratio relative to $BR(\Phi\to gg)$ (see
 \leqn{BRRdef}) for a 5
  $\sigma$ observation with a 900~fb$^{-1}$ data set as described in
  the text.}
\end{center}
\end{table}

To conclude, we list aspects of the PLC measurements of $\Phi$ that
would 
clearly advance our knowledge over what will be available from LHC:
\begin{itemize}
\item  The PLC will measure the $\Phi$ branching fractions to decay
  modes difficult to access at the LHC and, most probably, will reveal
  new decay modes that are not visible at the LHC above background.
\item Although we can infer the value of the $\Gamma(\Phi\to gg)$ from
  the assumption that the the production at the LHC is dominated by
  $gg\to \Phi$, it is very difficult to check this assumption directly
  from LHC measurements.  At the PLC, the quantity \leqn{refrat} can
  be measured from a known $\gamma\gamma$ initial state.  If this
  measurement agrees with the LHC value, this will confirm the
  production process assumed in the analysis of LHC measurements. 
\item Though the spin of the $\Phi$ will be measured from the angular
  distribution of $\Phi$ decay products at the LHC, the PLC will
  provide a sharp test of the spin assignment:   A spin 0 $\Phi$ will
  be produced only from the $J^3 = 0$ state of two photons; a spin 2
  $\Phi$ will be produced from the $J^3 = \pm 2$ states.
 \item The CP of the $\Phi$ can be measured directly  using 
  transversely polarized initial photons, by comparing the production
  cross section for parallel and perpendicular polarizations.
\item  Since significant limits can be placed on all possible 2-body
  decay modes of the $\Phi$, the PLC will allow us estimate the total rate
  of $\Phi$ production and the absolute value of the $\Phi$ width.
\end{itemize}

\subsection{Resonant production $\ee \to \Phi$}

If the $\Phi$ has spin 2, the  coupling to electron pairs is not
necessarily helicity-suppressed.   This opens the possibility of
resonant production in $\ee$ collisions.    This possibility is
realized in some of the models discussed in Section 3.6. Indeed, the
presence or absence of a $\Phi$ coupling to $\ee$ is a crucial
diagnostic of those models, so it is important to obtain either an
observation of the resonance or a very strong limit on its production.

For a narrow resonance and for unpolarized beams with the
center-of-mass energy $\sqrt{s}$,
 the total cross section can be expressed as 
\beq
\sigma(s) = 4\pi^2 (2 J + 1) {\Gamma (\Phi \to e^+ e^- ) \over m_\Phi}
\delta (s - m_\Phi^2) \ , 
\eeqn
where $J$ is the spin of the resonance.
Assuming a narrow resonance and a Gaussian beam energy distribution
centered around  $750$~GeV
 with the width $\sigma_M$:
\beq
\sigma (e^+e^- \to \Phi) = 2\pi^2 (2 J + 1)
 {\Gamma (\Phi \to e^+ e^- ) \over \sqrt{2 \pi} m_\Phi^2\sigma_M}. 
\eeqn
A partial width  $\Gamma (\Phi \to e^+ e^- ) = 0.1$~MeV  is consistent
with the LHC observations. 
Assuming  $\sigma_M/m_\Phi \approx 1\%$, one then finds  
$\sigma (e^+e^- \to \Phi)  \approx 0.4$~pb. 
With the  integrated luminosity of
$8$~ab$^{-1}$~\cite{Barklow:2015tja}, 
this implies
of order $3\times 10^{6}$ signal events.   Thus it is possible
either to study the full range of $\Phi$ decay modes in $\ee$
collisions or to put a strongly constraining limit on the coupling of
$\Phi$ to $\ee$.

\begin{table}
\begin{center} 
\begin{tabular}{l|cccccccc}

&  $qq+gg$ &  $b \bar b$ & $t \bar t$ &$\tau\tau$& 
$\mu \mu$ & $\gamma \gamma$  & $WW$ &  $ZZ $  
\\    \hline 
$\sigma_{bkgd}$ (fb)&  1000 & 320  &  300 &  200 & 200 & 250 & 400 & 50 
 \\   
$\sigma_{5\sigma}$ (fb)  &2&  1 & 1 &0.8  & 0.8 & 0.9 & 1.1& 0.4
  \\    
\end{tabular}
\caption{ \label{tab:eeback}   Standard Model background cross
  sections for the observation of $\Phi$ as a resonance in $\ee$
  annihilation to the given final states at the ILC.  The second
  line gives an estimate of the cross section  for a 5
  $\sigma$ observation with a 8~ab$^{-1}$ data set.  This is to be
  compared to the expected cross section of 400 fb for observation in
  the $gg$ final states
  if $\Gamma(\Phi\to \ee)$ is just below the current experimental limit.}
\end{center}
\end{table}

The signal and background for different decay channels of the 
KK graviton are summarized in Table~\ref{tab:eeback}.
For $\gamma \gamma$, $WW$, and $ZZ$ channels the background cross
section is quoted after the cut  $|\cos \theta | < 0.8$; the remaining
background cross sections are inclusive.
For each decay channel, we estimate the cross section for 5 $\sigma$
observation as  $S/\sqrt{B}$. 

Assuming conservatively that the
dominant decay mode of the $\Phi$ is $gg$, the $\Phi$ would be
observed
as a resonance in $\ee\to gg$.  Then the presence of a resonance
could be observed at a factor 200 below the
current limit from LHC Drell-Yan data, corresponding to a $\Phi\to
\ee$ branching ratio of $10^{-4}$.  This is superior to
the expectation for the High-Luminosity LHC, for which we expect an
improvement of roughly $\sqrt{N} = 30$ over the current limit.  Note
that the sensitivity is even  higher in other final states, if one of these
turns out to be the dominant decay mode of the $\Phi$. 

If the resonance is observed above background, the 
$\ee$ observation offers additional opportunities.
 The reach in cross section is relatively
independent of the decay channel, so 
the complete phenomenological picture of $\Phi $ decays can be
assembled.  Invisible decays of the $\Phi$ could be seen down to
similar cross section levels using the process 
$\ee\to \Phi + \gamma$~\cite{Richard:2016nhm},
which is dramatically enhanced, in this scenario, over the values
 presented in Section 4.1.

Thus, if the LHC were to confirm the spin 2 nature of the $\Phi$
resonance, and if especially if a resonance found at the LHC or
precision measurements at the ILC provided evidence for the
presence of RS excitations above the TeV scale, the 1 TeV upgraded ILC
would be able to make unique measurements of the gravity sector of the
RS model.

\section{Conclusion}

In this report, we have reviewed the implications that the discovery
of a $\gamma\gamma$ resonance at 750~GeV would have on physics at 
the International Linear Collider.  This resonance has been
taken seriously by the theory community, at least to the extent that
a very large number of theoretical proposals for the identity of the
resonance have been put forward.  The $\gamma\gamma$ thus provides an
interesting case study on the implications of a discovery at the LHC
for trhe program of the ILC.  To carry out this study, we have surveyed these
theoretical proposals in terms of  their predictions for precision
observables that will be measured at the ILC.  We have also made
estimates for the direct production of the 750~GeV particle at a
second-stage ILC upgraded to an energy of 1~TeV.  These estimates
demonstrate
 that,
for reasonable parameter choices, 
the full set of decay channels of the resonance can be observed in
$\gamma\gamma$ and  possibly also in $\ee$ collisions.

It is already well understood that precision measurements of the
properties of the Higgs boson, the top quark, and $W$ boson, and
fermion-scattering will reveal a wealth of information on physics at
the TeV mass scale, providing a window to new physics complementary to
the one that will be provided by particle searches at the LHC.  In
this report, we have shown that the value of these measurements is
made even more clear with the example of the 750~GeV resonance.  The
interpretations put forward for this resonance span a very wide range
of models.   It is always possible that the new particle exists in its
own sector and has no  relation to the known Standard Model
particles.  However, most models of the resonance require a specific
role for this particle within models of electroweak symmetry
breaking, dark matter, and other effects expected at the TeV mass
scale.   Through these relations, the new resonance and other new
particles that must be associated with it will leave a characteristic 
 imprint in the
precision observables that would be observed at the ILC.   In this
way, the ILC measurements will test and dramatically winnow these
models.   While we hope for the discovery of additional new particles
at the LHC that will shed further light on the nature of the new resonance,
the information from these tests at the ILC will be available whether
or not the additional particles are within the energy reach of the LHC.

The conclusions of this report have an important implication for the 
choice of the next frontier accelerator in high-energy physics.   In
discussions of the implications of the 750~GeV resonance, we 
often hear as a first reaction that it
implies a need to construct a proton-proton accelerator of  higher
energy to discover other new particles associated with this resonance.
   Of course, higher  energy
is always desireable.    But there is no mature  technology
available today to raise the  energy of proton-proton
collisions.  More importantly, the models for the
resonance that we have reviewed in this report do not give a clear
goal for the masses of  these new particles and the specific collider energy
that would be needed to reach them.

In contrast to this,  precision measurements from the ILC will give
us  qualitative information that will narrow the diversity of models and 
provide  insight into the nature of the new physics that the
750~GeV resonance implies must be present.  And, the ILC can be 
constructed today.   We ought to build the
ILC and benefit from  the insight that 
it will give us as we plan for the longer term
future of high-energy physics.

\Acknowledgements

MEP, TB, and YG thank the participants in the Kavli Institute for Theoretical
Physics
 program ``New Accelerators for
the 21st Century'', and especially Nathaniel Craig, Joseph Incandela,
and Liantao Wang, for useful discussions of the issues presented here.

\end{document}